\documentclass[12pt,preprint]{aastex}

\slugcomment{Astrophysical Journal}

\shorttitle{Nuclear starbursts in Seyfert 1s and 2s}
\shortauthors{Imanishi \& Wada}

\begin{document}


\title{Comparison of Nuclear Starburst Luminosities between Seyfert
1 and 2 Galaxies Based on Near-infrared Spectroscopy}


\author{Masatoshi Imanishi\altaffilmark{1,2} and Keiichi
Wada\altaffilmark{1} }
\affil{National Astronomical Observatory, 2-21-1, Osawa, Mitaka, Tokyo
181-8588, Japan}
\email{imanishi@optik.mtk.nao.ac.jp, wada.keiichi@nao.ac.jp}

\altaffiltext{1}{Visiting Astronomer at the Infrared Telescope Facility,
which is operated by the University of Hawaii under
Cooperative Agreement no. NCC 5-538 with the National
Aeronautics and Space Administration, Office of Space
Science, Planetary Astronomy Program.}

\altaffiltext{2}{Based in part on data collected at Subaru Telescope,
which is operated by the National Astronomical Observatory of Japan.}


\begin{abstract}
We report on infrared $K$- (2--2.5 $\mu$m) and $L$-band (2.8--4.1
$\mu$m) slit spectroscopy of 23 Seyfert 1 galaxies in the CfA and
12 $\mu$m samples.
A polycyclic aromatic hydrocarbon (PAH) emission feature at 3.3 $\mu$m in
the $L$ band is primarily used to investigate nuclear star-forming
activity in these galaxies.
The 3.3 $\mu$m PAH emission is detected in 10 sources (=43\%),
demonstrating that detection of nuclear star-formation in a significant
fraction of Seyfert 1 galaxies is now feasible.
For the PAH-detected nuclei, the surface brightness values of the PAH
emission
are as high as those of typical starbursts, suggesting
that the PAH emission probes the putative nuclear starbursts in the
dusty tori around the central active galactic nuclei (AGNs). 
The magnitudes of the nuclear starbursts are quantitatively estimated
from the observed 3.3 $\mu$m PAH emission luminosities.
The estimated starburst luminosities relative to some indicators of AGN
powers in these Seyfert 1s are compared with 32 Seyfert 2s in the same
samples that we have previously observed. 
We find that there is no significant difference in 
nuclear starburst to AGN luminosity ratios between Seyfert 1 and 2
galaxies, and that nuclear starburst luminosity positively
correlates with AGN power in both types of Seyferts. 
Our results favor a {\it slightly-modified} AGN unification model, which 
predicts that nuclear starbursts occurring in the
dusty tori of Seyferts are physically connected to
the central AGNs, rather than the {\it classical} unification paradigm,
in which the dusty tori simply hide the central AGNs of Seyfert 2s and 
reprocess AGN radiation as infrared dust emission in Seyferts. 
No significant differences in nuclear star formation properties are 
recognizable between Seyfert 1s in the CfA and 12 $\mu$m samples.
\end{abstract}


\section{Introduction}

Seyfert galaxies constitute a population that shows bright optical
nuclei \citep{sey43}, and the most numerous class of active galactic 
nuclei (AGNs) in the local universe.
It is generally believed that they contain a mass-accreting supermassive
black hole as a central engine, which is surrounded by toroidally
distributed dust and molecular gas (the so-called ``dusty torus''). 
There are two types of Seyfert galaxy, type-1 (which show broad
optical emission lines) and type-2 (which do not).
The different emission properties are explained by the AGN unification 
paradigm, in which the central engine of a Seyfert 2 galaxy is
intrinsically the same as that in a Seyfert 1 galaxy, but is obscured
from our line-of-sight by a dusty torus \citep{ant93}.
In this {\it classical} AGN unification paradigm, intrinsic torus
properties are assumed to be the same for the two types of Seyfert galaxy,
and the primary role of the dusty tori is twofold: they simply (1) hide
the central AGNs in Seyfert 2 galaxies, and (2) absorb AGN radiation and
re-radiate as infrared dust emission in both Seyfert 1 and 2 galaxies
\citep{pie92}.  

Since a dusty torus is rich in molecular gas, {\it nuclear} starbursts
may occur there, and energy feedback from the nuclear starbursts could
work to inflate the torus \citep{fab98}. 
Using high-resolution three-dimensional hydrodynamic simulations, 
\citet{wad02} showed that the nuclear starbursts in an extended (tens of 
pc) dusty torus can produce an inflated, turbulent torus around
a central supermassive black hole. 
They also suggested the possibility that the nuclear starbursts in the
torus may be more powerful in Seyfert 2s than in Seyfert 1s, because 
in a more inflated torus with stronger nuclear starbursts, the chance 
that the central engine is obscured and that, therefore, the galaxy would
be classified
as a Seyfert 2, would be higher. 
From a study of the equivalent width of the iron K$\alpha$ emission at
6.4 keV in X-rays, \citet{lev02} made a similar suggestion.
To obtain a deeper understanding of AGNs, it is important to
observationally constrain the putative nuclear starbursts in the dusty 
tori of Seyferts, and to test whether this {\it slightly-modified} AGN
model (with nuclear starbursts in the dusty tori) is more representative
of actual AGNs than the {\it classical} model.

Some observations and simple theoretical considerations of gravitational
instability suggest that nuclear starbursts in Seyfert galaxies are
likely to occur at the outer part of the dusty torus \citep{hec97,ima03}.
Attempts have been made to detect nuclear starbursts in the tori of
Seyfert galaxies by
several groups in the UV--optical 
\citep{gon01} and near-infrared at $\lambda$ $<$ 2.5 $\mu$m \citep{oli99}.
Detection has been successful only in type-2 sources, 
because unattenuated AGN emission from Seyfert 1 galaxies
weakens the signatures of the nuclear starbursts in the observed spectra and
makes their detection difficult; however, it is also possible that 
Seyfert 1s have weaker nuclear starbursts than do Seyfert 2s
\citep{wad02}.  
Until very recently, very little about nuclear starbursts in Seyfert
1 galaxies has been constrained observationally.

To study nuclear starbursts in Seyfert galaxies, infrared $L$-band
(2.8--4.1 $\mu$m) spectroscopy can be a very powerful tool, as stated by
\citet{ima02}.
First, a starburst usually shows a strong polycyclic aromatic
hydrocarbon (PAH) emission feature at 3.3 $\mu$m, whereas a pure AGN shows
only a featureless continuum, with virtually no PAH emission.
Thus, from an $L$-band spectrum, we can estimate the fractional
contribution from a starburst and an AGN to the observed flux
\citep{imd00}.
Second, the 3.3 $\mu$m PAH emission feature is intrinsically so
strong that even a weak starburst is detectable \citep{ima02}.
In fact, weak starbursts that had previously been undetected with
other methods have been newly found in some Seyfert 2 galaxies
\citep{ima02}.
Third, dust extinction is much lower in the $L$-band ($A_{\rm L}$/$A_{\rm V}$
$\sim$ 0.06; Rieke \& Lebofsky 1985) than at shorter
wavelengths, so that differences in the bright AGN glare emission
of Seyfert 1s and 2s caused by the flux attenuation by the dusty torus
are significantly reduced, as compared to shorter wavelengths.
This suggests that, given the success of the detection of nuclear
starbursts in a significant fraction of Seyfert 2 galaxies
\citep{ima02,ima03}, similar detection may also be feasible
in many Seyfert 1 galaxies. 
This expectation has actually been confirmed by the successful
detection of nuclear starbursts in a few type-1 Seyferts
\citep{rod03}.
Finally, it has been proven that nuclear starbursts in the tori of
Seyfert galaxies are reasonably quantifiable from 
observed 3.3 $\mu$m PAH emission luminosities, owing mainly to the low
dust extinction in the $L$-band \citep{ima02}.
All of these advantages make $L$-band spectroscopy an excellent tool to
quantitatively determine nuclear starburst luminosities in 
Seyfert galaxies, and to compare the starburst luminosities
in Seyfert 1 and 2 galaxies.    

Besides $L$-band spectra, $K$-band (2--2.5 $\mu$m) spectra can also be
used to study stellar emission in Seyfert nuclei.
Since the CO absorption features at $\lambda_{\rm rest}$ = 2.3--2.4
$\mu$m in the rest-frame are produced by stars, and not by a pure AGN,
emission from stars and AGNs is distinguishable using these
features \citep{oli99,iva00,ia04}. 
Furthermore, nuclear $K-L$ colors provide useful information on the
origin of nuclear emission from Seyfert galaxies \citep{alo03,wil84}.
A combination of $L$- and $K$-band spectra can add further constraints
on the nuclear emission properties in Seyfert galaxies.

We have extensively studied nuclear starbursts in Seyfert 2
galaxies through $L$- and $K$-band spectroscopy \citep{ima03,ia04}.
In this paper, we report the results of $L$- and $K$-band
spectroscopy of Seyfert 1 galaxies, to compare with the data for
Seyfert 2 galaxies.
Throughout the paper, $H_{0}$ $=$ 75 km s$^{-1}$ Mpc$^{-1}$,
$\Omega_{\rm M}$ = 0.3, and $\Omega_{\rm \Lambda}$ = 0.7 are adopted.

\section{Targets}

The Seyfert 2 galaxies previously studied by \citet{ima03} and
\citet{ia04} were taken from the CfA \citep{huc92} and 12 $\mu$m (Rush,
Malkan, \& Spinoglio 1993) samples.
The CfA and 12 $\mu$m samples were selected through optical spectroscopy
of large numbers of galaxies limited by optical and {\it IRAS} 12 $\mu$m
fluxes, respectively, and are thought not to be strongly biased toward
or against the presence of nuclear starbursts in the torus. 
Our aim is to investigate the emission properties of the {\it nuclear}
starbursts in the torus, rather than more powerful {\it circumnuclear}
starbursts in host galaxies, which are often ring-shaped at $\sim$kpc
scale from the center \citep{sto96}.
Since we make the investigation through ground-based spectroscopy using
a 1--2 arcsec wide slit, Seyfert 1 galaxies at $z$=0.008--0.035 are
selected as the targets in the same way as the Seyfert 2
galaxies used for comparison (see Imanishi 2003), where 1
arcsec corresponds to a physical scale of 150 pc ($z$ =
0.008) to 650 pc ($z$ = 0.035). 
To be best observable from Mauna Kea, Hawaii (our observing
site), the declinations of Seyfert 1 galaxies are limited to
being larger than $-$30$^{\circ}$.
Owing to the telescope limit of the IRTF 3 m telescope that was mainly
used in this study, a restriction of declination of $<$68$^{\circ}$
is also applied. 
These selection criteria result in 14 and 22 Seyfert 1 galaxies in the
CfA and 12 $\mu$m samples, respectively.
Of these, six sources are included in both samples, totaling 30
Seyfert 1 galaxies.
Two Seyfert 1 galaxies, NGC 4253 and Mrk 1239, had $L$-band
spectra available at the time of our observations \citep{rod03}.
Five Seyfert 1 galaxies (NGC 3080, Mrk 789, NGC 6104, MCG-3-7-11, and NGC
4602) could not be observed owing to our limited telescope time.
We observed the remaining 23 sources.
Detailed information on the observed 25 Seyfert 1 galaxies (23 sources by
ourselves and 2 sources by Rodriguez-Ardila \& Viegas) is
summarized in Table 1.
The infrared luminosities of the majority of them are
$<$10$^{11}$L$_{\odot}$, with the highest one with
$\sim$10$^{11.5}$L$_{\odot}$ (NGC 7469), so that the observed targets
are moderately infrared luminous Seyfert 1s, with no ultraluminous
infrared galaxies (L$_{\rm IR}$ $>$ 10$^{12}$L$_{\odot}$; Sanders \&
Mirabel 1996) included. 
There are no obvious biases in the 25 objects selected, and so the
observed sample is large enough and relatively unbiased such as to be
nearly statistically complete. 
Thus, these datasets provide important information on nuclear starbursts
in Seyfert 1 galaxies.

\section{Observations}

All observations were made with the SpeX \citep{ray03} at the IRTF 3 m
telescope on Mauna Kea, Hawaii, with the exception of NGC 7469, which was
observed with IRCS \citep{kob00} at the Subaru 8.2 m telescope at Mauna Kea.
Table 2 tabulates the detailed information on these observing runs.

For the IRTF SpeX observing runs, the 1.9--4.2 $\mu$m cross-dispersed mode
with a 1\farcs6 wide slit was employed.
This mode enables $L$- (2.8--4.1 $\mu$m) and $K$-band (2--2.5
$\mu$m) spectra to be obtained simultaneously, with a
spectral resolution of R $\sim$ 500. 
The sky conditions were photometric throughout the observations, and the
seeing sizes at $K$ were in the range 0$\farcs$4--1$\farcs$0 (full-width
at half-maximum; FWHM).
The position angle of the slit was set along the north-south direction.
A standard telescope nodding technique (ABBA pattern) with a throw of 7.5
arcsec was employed along the slit to subtract background emission.
The telescope tracking was monitored with the infrared slit-viewer of SpeX.
Each exposure was 15 sec, and 2 coadds were made at each position.

The Subaru IRCS observation of NGC 7469 was made using a 0$\farcs$9-wide
slit, and with a 58-mas pixel scale.
The achievable spectral resolution at 3.5 $\mu$m was R $\sim$ 140.
The seeing at $K$ was $\sim$0$\farcs$7 in FWHM. 
The weather was not photometric, owing to the presence of cirrus.
A standard telescope nodding technique (ABBA pattern), with a throw of 7
arcsec along the slit, was employed.
Each exposure was 2.5 sec, and 15 coadds were made at each position.

F- and G-type main sequence stars (Table~\ref{tbl-2}) were observed as
standard stars, with airmass difference of $<$0.1 to the individual
Seyfert 1 nuclei, to correct for the transmission of the Earth's
atmosphere.
The $K$- and $L$-band magnitudes of the standard stars were estimated
from their $V$-band (0.6 $\mu$m) magnitudes, adopting the $V-K$ and
$V-L$ colors appropriate to the stellar types of individual standard
stars, respectively \citep{tok00}.

Standard data reduction procedures were employed using IRAF
\footnote{
IRAF is distributed by the National Optical Astronomy Observatories,
operated by the Association of Universities for Research
in Astronomy, Inc. (AURA), under cooperative agreement with the
National Science Foundation.},
in the same way as employed for Seyfert 2 galaxies \citep{ima03}.
Initially, bad pixels and pixels hit by cosmic rays were replaced with the
interpolated values of the surrounding pixels.
Then, frames taken with an A (or B) beam were subtracted from
frames subsequently taken with a B (or A) beam, and
the resulting subtracted frames were added and divided by a
spectroscopic flat image.
The spectral extraction of Seyfert 1 galaxies and standard stars 
along the slit was then
made by integrating signals over 2$\farcs$1--3$\farcs$0 for the
IRTF SpeX data and 1$\farcs$1 for the Subaru IRCS data.
Wavelength calibration was performed using the wavelength-dependent
transmission of the Earth's atmosphere.
The Seyfert 1 spectra were divided by those of the corresponding standard
stars, and were multiplied by the spectra of blackbodies with
temperatures corresponding to those of the individual standard stars
(Table~\ref{tbl-2}).
Appropriate spectral binning was applied for faint sources.
Flux calibration was made using signals detected inside our slit spectra,
with the exception of NGC 7469, the flux calibration of
which was made using a previous photometric measurement ($L$
= 8.36 mag; 1$\farcs$4 aperture) by \citet{mar98}.

\section{Results}

\subsection{$L$-band}

Figure 1 presents flux-calibrated $L$-band slit spectra of the 23 Seyfert
1 nuclei.
Since the nuclear $L$-band emission from Seyfert 1 galaxies is often
dominated by compact, spatially-unresolved emission \citep{alo98,alo03},
nuclear $L$-band fluxes measured with slightly different apertures
should, in principle, be similar.
In fact, flux levels based on our slit spectra are in reasonable agreement
(mostly $<$0.5 mag) with previous photometric measurements using $<$5
arcsec apertures, suggesting that possible flux loss in our slit spectra
will not significantly affect emission flux estimates.

Many sources show clear excess at the expected wavelength of the
3.3 $\mu$m PAH emission feature at (1 + $z$) $\times$ 3.29 $\mu$m.
When at least two successive data points are significantly higher than
an adopted continuum level, we regard the 3.3 $\mu$m PAH emission
feature as detected.
Using this criterion, 3.3 $\mu$m PAH emission is detected in 10 Seyfert
1 nuclei, marked with ``3.3 $\mu$m PAH'' in Fig. 1.
The fluxes, luminosities, and rest-frame equivalent widths
(EW$_{\rm 3.3PAH}$) of the 3.3 $\mu$m PAH emission are estimated using
the method described by \citet{ima02} and are summarized in Table~\ref{tbl-3}.

Although PAH emission from nuclear star-formation is detected in
$\sim$43\% (=10/23) of the observed Seyfert 1 nuclei, the significantly
smaller EW$_{\rm 3.3PAH}$ values, as compared to star-formation ($\sim$100 nm;
Imanishi \& Dudley 2000), suggest that a featureless continuum from hot
dust heated by AGNs contributes dominantly to the observed $L$-band
fluxes, and dilutes the 3.3 $\mu$m PAH emission.
To estimate the energetic importance of the detected nuclear
star formation to the total luminosities of these Seyfert 1 galaxies, 
we investigate the nuclear 3.3 $\mu$m PAH to infrared luminosity ratios 
(L$_{\rm nuclear-3.3PAH}$/L$_{\rm IR}$), as we did for Seyfert 2 galaxies 
\citep{ima02,ima03}.
The L$_{\rm nuclear-3.3PAH}$/L$_{\rm IR}$ ratios in these Seyfert 1
galaxies are summarized in column 5 of Table 3.  
As was the case for Seyfert 2 galaxies in the CfA and 12 $\mu$m samples 
\citep{ima03}, the ratios in Seyfert 1 galaxies are also much smaller
than those of star-formation dominated galaxies ($\sim$10$^{-3}$; Mouri
et al. 1990; Imanishi 2002).
This suggests that the detected nuclear star-formation contributes
little to the total infrared luminosities of these Seyfert 1 galaxies,
which should be dominated by AGNs and/or extended star-formation in host
galaxies.   

\subsection{$K$-band}

For sources observed with the IRTF SpeX, $K$-band spectra are also
available.
Figure 2 shows flux-calibrated $K$-band slit spectra of 22 sources,
with the exception of NGC 7469.

\subsubsection{$K - L$ Colors}

Since the SpeX $K$- and $L$-band spectra were obtained simultaneously,
possible slit loss should be at a similar level.
Thus, $K - L$ colors should be less uncertain than the absolute $K$- or
$L$-band magnitudes themselves.
Table 4 summarizes the nuclear $K - L$ colors measured with our slit
spectra.

The majority of the observed Seyfert 1 galaxies show $K-L$= 1--2 mag,
which is similar to the intrinsic AGN colors of Seyfert 1 galaxies
\citep{alo03}, but much redder than those of star-formation ($K-L <$
0.4; Willner et al. 1984).
This suggests that the $K$- and $L$-band emission from Seyfert 1
nuclei detected inside our slit spectra come largely from AGN-originated
emission.

\subsubsection{Emission Lines}

The Br$\gamma$ ($\lambda_{\rm rest}$ = 2.166 $\mu$m) emission
lines are discernible in the spectra of several sources.
The line widths of the Br$\gamma$ emission in Seyfert 1 galaxies are so
broad that the Br$\gamma$ fluxes are reasonably quantifiable from our
moderate-resolution (R $\sim$ 500) spectra.
For Seyfert 1 galaxies with clear Br$\gamma$ detection, we fit the
Br$\gamma$ emission lines with Gaussian profiles, which are shown as
solid lines in Figure 2. 
The derived FWHMs, fluxes, luminosities, and rest-frame equivalent
widths of the Br$\gamma$ emission lines are summarized in Table 5.

Besides the Br$\gamma$ emission lines, there are possible
signs of narrow ($<$500 km s$^{-1}$ in FWHM) emission lines,
such as H$_{2}$ 1--0 S(1) at $\lambda_{\rm rest}$ = 2.12 $\mu$m, in some
sources. 
To discuss these narrow emission lines in a quantitatively reliable
manner, by tracing their profiles, spectral-resolution of R $>$ 1000 is
desirable.
No quantitative estimates for narrow emission lines are made in our
lower resolution (R $\sim$ 500) spectra.

\subsubsection{CO Absorption Features}

Some sources show flux depression at $\lambda_{\rm rest}$ $>$ 2.3
$\mu$m, as compared to shorter wavelengths.
This is usually attributed to CO absorption at 2.3--2.4 $\mu$m,
caused by stars.
To estimate its strength, we adopt the spectroscopic CO index proposed by
Doyon, Joseph, \& Wright (1994).
This index is defined as
\begin{eqnarray}
{\rm CO_{spec}} & \equiv & -2.5log<R_{2.36}>,
\end{eqnarray}
where $<$R$_{2.36}$$>$ is the average of actual signals at
$\lambda_{\rm rest}$ = 2.31--2.40 $\mu$m divided by a power-law
continuum (F$_{\rm \lambda}$ = $\alpha \times \lambda^{\beta}$)
extrapolated from shorter wavelengths.
In our spectra, data points at $\lambda_{\rm rest}$ = 2.1--2.29 $\mu$m,
excluding clear emission lines, are used to determine a continuum
level, because this wavelength range is fully covered in the
$\lambda_{\rm obs}$ $=$ 2.07--2.5 $\mu$m (observed frame) spectra.
Although this wavelength range is slightly different from that employed
by Doyon et al. ($\lambda_{\rm rest}$ = 2.0--2.29 $\mu$m), our
spectroscopic CO index is essentially the same as theirs.
For sources with clear CO absorption signatures,
the adopted continuum levels are shown as dashed lines in Figure 2 and
the derived CO indices are summarized in Table 6.

Even for the detected sources, with the exception of NGC 2639, the CO
absorption feature is weak.
The major uncertainty of the CO index comes from the ambiguity of
the continuum determination, which is difficult to assess quantitatively.
For other sources without signatures of the absorption, we simply regard
the CO absorption as undetected.

Correction of the observed spectroscopic CO index for the dilution by
AGNs' featureless continuum in the $K$-band could, in principle,
enable us to investigate nuclear star-formation properties in more detail,
as has been undertaken in Seyfert 2 galaxies \citep{iva00,ia04}.
However, for this correction process to work reasonably well,
the observed spectroscopic CO index must be $>$0.1.
In Seyfert 1 galaxies, the observed CO absorption feature is usually much
shallower than in Seyfert 2 galaxies, owing to a higher contribution from
unattenuated AGNs' featureless continuum in the former.
Thus, no correction of the observed spectroscopic CO index for AGNs'
dilution is attempted.

NGC 2639 is an exception; its observed spectroscopic CO index,
without any AGN correction, is as large as those found in many
star-forming galaxies \citep{coz01}.
The $K - L$ color of this source is unusually blue as compared to the other
Seyfert 1 galaxies in our sample (Table 4).
The contribution from nuclear star-formation to the observed $K$-band
flux must be high in this Seyfert 1 galaxy.

\section{Discussion}

\subsection{Origins of the Detected Stellar Emission}

The nuclear $K$- and $L$-band fluxes from Seyfert 1 galaxies are the
superposition of emission from (1) the AGN, (2) putative nuclear
starbursts in the dusty torus, (3) the central part of old bulge stars, and
(4) the central part of disk stars.
The 3.3 $\mu$m PAH emission can, in principle, originate in the latter
three components.
Nuclear starbursts and quiescent bulge and disk star-formation
can be distinguished, based on the emission surface brightness of
star-forming activity.

\citet{hec01} estimated the star-formation rate per unit area of
10$^{-1}$ M$_{\odot}$ yr$^{-1}$ kpc$^{-2}$ as a lower limit, beyond which
the superwind activity, a characteristic property of starbursts, can
occur.
This limit corresponds to the infrared surface brightness of
S$_{\rm IR}$ $\sim$ 2 $\times$ 10$^{42}$ ergs s$^{-1}$ kpc$^{-2}$
\citep{ken98}, or to the 3.3 $\mu$m PAH emission surface brightness of
S$_{\rm 3.3 PAH}$ $\sim$ 2 $\times$ 10$^{39}$ ergs s$^{-1}$ kpc$^{-2}$
\citep{ima02}.
The 3.3 $\mu$m PAH surface brightness values in the observed Seyfert 1
galaxies are summarized in the last column of Table 3.
At least for the PAH-detected Seyfert 1 nuclei, the surface brightness values 
of the 3.3 $\mu$m PAH emission are as high as those typically found in
starbursts.
We therefore conclude that the detected 3.3 $\mu$m PAH emission probes
putative nuclear starbursts in the dusty torus, rather than
quiescent star-formation.
For the PAH-undetected sources, the upper limits of the surface
brightness are also substantially above the threshold.

For Seyfert 2 galaxies in the CfA and 12 $\mu$m samples, \citet{ia04}
drew the same conclusion, based on the same calculation, that the detected 3.3 
$\mu$m PAH emission should come from nuclear starbursts in the dusty
tori.
Thus, we can use the 3.3 $\mu$m PAH emission as a useful probe for 
nuclear starbursts in both Seyfert 1 and 2 galaxies.

\subsection{Comparison of Nuclear Starburst Luminosities between Seyfert
1s and 2s}

\citet{ima02,ima03} showed that the luminosities of the nuclear
starbursts in {\it Seyfert 2s} are reasonably quantifiable from the
{\it observed} 3.3 $\mu$m PAH emission luminosities inside slit spectra,
primarily because (1) dust extinction is low at 3--4 $\mu$m, and (2) the
nuclear starbursts are thought to occur at the outer part of the dusty
torus, where obscuration is expected not to be severe.
Thus, we can make the reasonable assumption that the nuclear starburst
luminosities in {\it Seyfert 1s} are also quantitatively determined,
with reasonable accuracy, from the observed 3.3 $\mu$m PAH emission
luminosities. 

To compare nuclear starburst luminosities between Seyfert 1s and 2s,
we normalize the luminosities with AGN powers.
In both the classical and slightly-modified AGN unification models,
the AGN-powered infrared emission should be more
highly-attenuated in Seyfert 2s  
than in Seyfert 1s, owing to higher obscuration by the
dusty tori in the former.
However, a comparison of infrared spectral energy distributions between
Seyfert 1s and 2s has suggested that the obscuring effects for the AGN
emission from moderately infrared luminous Seyfert 2s, except the most
infrared luminous (L$_{\rm IR}$ $>$ 10$^{12}$L$_{\odot}$) and dustiest ones
(i.e., ultraluminous infrared galaxies), becomes insignificant at $>$10
$\mu$m \citep{alo03}.  
Since no Seyfert 2s studied by \citet{ima03} have 
L$_{\rm IR}$ $>$ 10$^{12}$L$_{\odot}$, we first use the {\it IRAS} 12
$\mu$m and 25 $\mu$m, and ground-based $N$-band (10.6 $\mu$m)
luminosities, because these luminosities have been argued to be 
good tracers of AGN powers in Seyfert galaxies, with less
contamination from host galaxies' emission than {\it IRAS} 60 $\mu$m and
100 $\mu$m luminosities \citep{spi89,rod97,alo02}.

Figures 3a, 3b, and 3c compare the {\it IRAS} 12 $\mu$m luminosities,
{\it IRAS} 25 $\mu$m luminosities, and $N$-band luminosities measured with
$>$4$''$ apertures using single-pixel detectors from the ground, with
the observed nuclear 3.3 $\mu$m PAH emission luminosities through our
slit spectroscopy.
Both Seyfert 1s (this paper) and 2s \citep{ima03} are plotted.
The abscissa and ordinate are taken to trace the powers of the AGNs and
nuclear starbursts, respectively \citep{ima02,ima03}.
If Seyfert 2s have intrinsically stronger nuclear starbursts in the
dusty tori, then their distribution should be to the upper-left side of
that of Seyfert 1s.
However, no significant difference is evident in the distribution of Seyfert
1s and 2s.
Both Seyfert 1s and 2s follow similar 3.3 $\mu$m PAH to 12 $\mu$m,
25 $\mu$m, and $N$-band luminosity ratios within the scatters.
The distribution of the absolute luminosities of the 3.3 $\mu$m PAH emission,
12 $\mu$m, 25 $\mu$m, and $N$-band is also similar in Seyfert 1s 
and 2s.
Even if there were some degree of obscuration effect for the AGN
emission from Seyfert 2s at $>$10 $\mu$m, correction would move the
plots of {\it Seyfert 2s to the right} in these figures.
In this case, the nuclear starburst to AGN luminosity ratios
in Seyfert 2s would be even smaller, which is opposite to the
predictions of \citet{wad02}.

In all the plots of Figs. 3a, 3b, and 3c, {\it IRAS} 12 $\mu$m,
{\it IRAS} 25 $\mu$m, and $N$-band aperture ($>$4$''$) measurements, which
were taken as good indicators of AGN powers, could contain emission from
star-formation in host galaxies, including the ring-shaped kpc-scale
circumnuclear starbursts, if they exist.
\citet{mai95} suggested that the host galaxies of Seyfert 2s are more
luminous than those of Seyfert 1s.
{\it IRAS} fluxes at longer wavelengths, such as 60 $\mu$m and
100 $\mu$m, may be severely contaminated by the host galaxies' emission.
However, the contamination is expected to be minimal at {\it IRAS} 12
$\mu$m and 25 $\mu$m \citep{spi89,rod97}.
If the host galaxies' contamination is significant, then correction for
the stronger contamination in Seyfert 2s will move the plots of Seyfert
2s to the left, as compared to Seyfert 1s, so that the plots
in Figs. 3a, 3b, and 3c is still consistent with the scenario of
stronger nuclear starbursts in Seyfert 2s \citep{wad02}. 

This possible ambiguity can be minimized if $N$-band photometry of
spatially-unresolved emission, measured with a small aperture ($<$2$''$),
is used to estimate the power of AGNs.
After the publication of \citet{ima03}, such photometric data have been
presented by \citet{gor04}.
These photometric measurements for Seyfert 1s in our sample are
summarized in column 9 of Table 1.
The same data for Seyfert 2s, studied by \citet{ima03}, together with
the estimated nuclear 3.3 $\mu$m PAH emission luminosities, are
summarized in Table 7.
Figure 3d compares the nuclear $N$-band luminosity measured with a
1$\farcs$5 aperture \citep{gor04} and nuclear 3.3 $\mu$m PAH emission
luminosities in Seyfert 1s and 2s.
As were the cases for Figs. 3a, 3b, and 3c, we see no clear evidence
that Seyfert 2s are distributed at substantially upper-left locations, as
compared to Seyfert 1s.

The ring-shaped kpc-scale circumnuclear starbursts correspond to 
1.5--7 arcsec for the observed Seyfert galaxies at $z$=0.008--0.035. 
Thus, if circumnuclear starbursts are important, the $N$-band
photometry measured with $>$4$''$ apertures should be systematically
higher than that with the 1$\farcs$5 aperture.
For the majority of the Seyfert 1s in Table 1, both measurements are
available, but they indicate that this is not the case.
In fact, although the circumnuclear starbursts were clearly detected in
the two nearby well-studied Seyferts, NGC 1068 and NGC 7469
\citep{mil96,lef01}, they were not commonly found in the $N$-band images
of the majority of Seyferts \citep{gor04}.
It is likely that possible time variation of the nuclear $N$-band
emission \citep{gor04}, rather than the inclusion or exclusion of the
circumnuclear starbursts, has more severe effects for the comparison of
the nuclear 3.3 $\mu$m PAH emission luminosities and ground-based $N$-band
luminosity measurements.
In this regard, the absence of a significant distribution change of
Seyfert 2s from Fig. 3c to Fig. 3d is reasonable.

Another powerful indicator of the power of AGNs is radio emission.
Radio emission from the core of a Seyfert galaxy can be taken as AGN
emission and suffers virtually no effects of extinction by the dusty
torus.
Core radio emission data, measured with 0$\farcs$25 spatial resolution,
are available for many of the CfA and 12 $\mu$m Seyferts  
\citep{kuk95,the00}.
These core radio data for Seyfert 1s and 2s are summarized in Table 1 
(column 10) and Table 7 (column 5), respectively.
Figure 3e plots the comparison between the core radio luminosities at 8.4 GHz
and nuclear 3.3 $\mu$m PAH emission luminosities in both Seyfert 1s and 2s.
Again, in this plot, we see no clear trend that Seyfert 2s are
distributed to the upper-left side of Seyfert 1s.

The 2--10 keV X-ray luminosity can also provide a good indication of the
power of an 
AGN, because an AGN usually produces 2--10 keV emission far more
strongly than does star-formation.
Some fraction of the Seyfert 1s and 2s in our sample have available
2--10 keV X-ray observations \citep{tur97,geo98}.
However, unlike previous AGN-power indicators where {\it observed}
luminosities have been used, {\it absorption-corrected} luminosity is
essential for 2--10 keV emission, particularly for Seyfert 2s.
This absorption-correction can be uncertain for many Seyfert 2s, because
as many as half of all Seyfert 2s are Compton thick 
(N$_{\rm H}$ $>$ 10$^{24}$ cm$^{-2}$; Risaliti et al. 1999), in which
case, observed 2--10 keV emission is dominated by a reflected/scattered
component, rather than by a direct transmitted component, making 
estimates of the absorption-corrected 2--10 keV luminosities highly
uncertain.
For Seyfert 1s, the absorption correction was thought to be more
straightforward and often insignificant, but recent
X-ray data of the Seyfert 1 galaxy, Mrk 231, imply that
the correction can be more complicated than has been anticipated
\citep{bra04}.
Additionally, the fraction of Seyferts in our sample with available
absorption-corrected 2--10 keV X-ray luminosities is smaller than that
with other previous AGN-power indicators.
No comparison between the 2--10 keV X-ray and nuclear 3.3 $\mu$m PAH
emission luminosities is made in this paper.

To summarize the above comparisons between the luminosities of the
nuclear starbursts and AGNs in Figs. 3a--3e, Seyfert 1s and 2s have
nuclear starbursts of similar strength, with respect to the AGN
luminosities.
We see no clear evidence that Seyfert 2s have systematically more
powerful nuclear starbursts than do Seyfert 1s.  

In Seyfert 2s, the correlation between nuclear 3.3 $\mu$m PAH emission
luminosities and {\it IRAS} 12 $\mu$m, {\it IRAS} 25 $\mu$m, and
ground-based $N$-band (10.6 $\mu$m) luminosities measured with $>$4$''$
apertures, was statistically confirmed \citep{ima03}, which suggests that
the powers of nuclear starbursts and AGNs are correlated.
We apply the generalized Kendall's rank correlation statistic
\citep{iso86}
\footnote{
software is available at:
http://www.astro.psu.edu/statcodes/.}
to both types of Seyferts in Figs 3a-3e.
The probability that a correlation is not present is found to be
0.01\%, $<$0.01\%, 0.05\%, 0.8\%, and 0.6\% for Fig. 3a, 3b, 3c, 3d,
and 3e, respectively.
Thus, the correlation between the luminosities of the nuclear starbursts
in the dusty tori and central AGNs is statistically confirmed in
Seyfert galaxies in all the plots in Fig. 3, suggesting the nuclear
starburst-AGN connection.
The enhancement of a mass accretion rate onto a central supermassive
black hole, owing to the increased turbulence of molecular gas in the torus
because of the presence of nuclear starbursts \citep{wad02}, is one possible
scenario that could account for the luminosity correlation.

Figures 4a and 4b show, respectively, the histograms of the redshifts
and infrared luminosities of the observed Seyfert 1s and 2s.
Although the fraction of high redshift sources is slightly higher in
Seyfert 1s than 2s, the distribution of the infrared luminosities is
similar.
Thus, the above conclusions cannot be caused by the different
luminosity distribution between the Seyfert 1 and 2 samples.
If Seyfert galaxies at higher redshifts had stronger nuclear
starbursts, relative to the AGN power, than those at lower redshifts,
our conclusions could be affected by some bias caused by the
slightly different redshift distribution between Seyfert 1s and 2s.
However, the redshift range probed by our study is only 0.008--0.035,
where no strong cosmological evolution is expected, and so this scenario
seems very unlikely.

Figures 5a and 5b show, respectively, the rest-frame equivalent widths of
the 3.3 $\mu$m PAH emission feature (EW$_{\rm 3.3PAH}$) and the
observed spectroscopic CO indices (CO$_{\rm spec}$) of the Seyfert 1s and
2s.
Both the EW$_{\rm 3.3 PAH}$ and CO$_{\rm spec}$ values are
systematically smaller in Seyfert 1s than in Seyfert 2s.
This trend is reasonably explained by the higher contributions from the
featureless AGN continuum to the near-infrared fluxes in Seyfert1s, owing
to less attenuation, which decrease both the EW$_{\rm 3.3PAH}$ and
CO$_{\rm spec}$ values, as compared to Seyfert 2s.
   
The overall results revealed through the present datasets favor the 
{\it slightly-modified} AGN unification theory, in which:  
(1) the central engines of both Seyfert 1s and 2s are intrinsically the
    same, but those of Seyfert 2s are obscured by dusty tori; 
(2) the dusty tori not only obscure and 
    re-radiate AGN light as infrared dust emission, but also 
    {\it contain nuclear starbursts in both Seyfert 1s and 2s of
    similar strength}; and   
(3) {\it the nuclear starbursts have close physical connections to the
    central AGNs}.  

The presence of the nuclear starbursts in the dusty torus has some
important implications for AGN research.
First, in the calculation of infrared dust emission from the torus, 
only radiation from the central AGN has usually been taken into account
as a heating energy source. 
Although the absolute magnitudes of the nuclear starbursts, estimated in
this paper and by \citet{ima03}, are not very powerful, in terms of the
infrared luminosity from a whole Seyfert galaxy, the inclusion of 
nuclear starbursts may help us to better understand emission from the
dusty torus of a Seyfert galaxy \citep{van03}. 
Next, theoretically, the nuclear starbursts can naturally occur in an
extended (tens pc) torus, but they are difficult to produce in a
compact (smaller than a few pc) torus (Wada \& Norman 2002; Wada et al. in
preparation). 
Our detection of nuclear starbursts may suggest that the extended
torus model is more plausible in Seyfert galaxies.
Finally, radiation from nuclear starbursts could affect the mass
accretion rate onto a central supermassive black hole \citep{ume98,ohs99}. 
So far, more powerful, but more distant, {\it circumnuclear} starbursts
have been considered as the primary radiation source \citep{ume97}, but
{\it nuclear} starbursts, even though less powerful, may have important
effects in this respect, because of their proximity to the central supermassive
black hole.   

\subsection{Do the CfA and 12 $\mu$m Seyfert 1 Galaxies Have
Different Nuclear Star Formation Properties?}

So far we have combined the CfA and 12 $\mu$m Seyfert 1s,
and conducted a comparison with Seyfert 2s in the same samples.  
It is of interest to compare nuclear star formation properties, by 
dividing Seyfert 1s into two separate samples, as was done for
Seyfert 2s by \citet{ia04}.
Figure 6 plots the distribution of CO$_{\rm spec}$, EW$_{\rm 3.3PAH}$,
and L$_{\rm nuclear-3.3PAH}$/L$_{\rm IR}$, by separating the Seyfert 1s into 
two samples.
In all the plots, no systematic differences can be seen between
Seyfert 1s in the CfA and 12 $\mu$m samples.

\citet{ia04} previously found no systematic difference in Seyfert
2s between CfA and 12 $\mu$m samples.
When combined with the absence of a difference in Seyfert 1s, we can
argue that the nuclear star formation properties of both Seyfert 1s and 2s
in the CfA sample are similar to those in the 12 $\mu$m sample.
In particular, the L$_{\rm nuclear-3.3PAH}$/L$_{\rm IR}$ ratios in
Fig. 6c trace the relative contribution from nuclear starbursts to the
infrared luminosity.
\citet{ho01} put forward the possibility that 12 $\mu$m Seyferts may
be biased to those with unusually elevated levels of nuclear star
formation, because nuclear star formation can be a strong 12 $\mu$m
emitter.
However, our studies provide no support for this.

\section{Summary}

We presented near-infrared $K$- and $L$-band spectroscopic results of
23 Seyfert 1 galaxies in CfA and 12 $\mu$m samples.
Our $L$-band spectroscopic method successfully detected 3.3 $\mu$m 
PAH emission, the signature of nuclear star formation, in a significant
fraction (=43\%) of the observed Seyfert 1 nuclei.
The high surface brightness values of the 3.3 $\mu$m PAH emission
suggested  
that the detected signatures are most likely to come from putative
nuclear starbursts in the dusty tori, rather than from quiescent
star-formation in the nuclear regions of these Seyfert 1 galaxies. 
The properties of the nuclear starbursts in Seyfert 1s were 
compared with those in Seyfert 2s; we found that Seyfert 1s
possess nuclear starbursts with similar luminosities to those of Seyfert 2s.
The nuclear starburst luminosities positively correlated with the AGN
luminosities in a statistical sense, suggesting nuclear
starburst-AGN connections in both Seyfert 1s and 2s.
No systematic differences in the nuclear star formation properties 
were found between Seyfert 1s in the CfA and 12 $\mu$m samples,
as previously found in Seyfert 2s; this suggested that 12 
$\mu$m Seyferts are not strongly biased to those with enhanced nuclear
star formation activity, as compared to the CfA sample.   

\acknowledgments

We are grateful to P. Sears, S. B. Bus (IRTF) and H. Terada, and
R. Potter (Subaru) for their supports during our observing runs.
The referee, D. Sanders, gave useful comments on the manuscript. 
KW is supported by Grant-in-Aids for Scientific Research (no. 15684003).
Some part of the data analysis was made using a computer system operated
by the Astronomical Data Analysis Center (ADAC) and the Subaru Telescope
of the National Astronomical Observatory, Japan.
This research has made use of the SIMBAD database, operated at CDS,
Strasbourg, France, and of the NASA/IPAC Extragalactic Database
(NED) which is operated by the Jet Propulsion Laboratory, California
Institute of Technology, under contract with the National Aeronautics
and Space Administration.

\clearpage

\begin{deluxetable}{lcrrrrcccccc}
\tabletypesize{\scriptsize}
\rotate
\tablecaption{Detailed Information on the Observed Seyfert 1 Galaxies
\label{tbl-1}}
\tablenum{1}
\tablewidth{0pt}
\tablehead{
\colhead{Object} & \colhead{Redshift}   & \colhead{f$_{\rm 12}$}   & 
\colhead{f$_{\rm 25}$}   & \colhead{f$_{\rm 60}$}   &
\colhead{f$_{\rm 100}$}  & 
\colhead{log L$_{\rm IR}$ (log L$_{\rm IR}$/L$_{\odot}$)} &
\colhead{f$_{\rm N}$} & \colhead{f$_{\rm N}$(1$\farcs$5)} & \colhead{log
f$_{8.4 \rm GHz}$} & \colhead{Remarks} \\
\colhead{} & \colhead{} & \colhead{(Jy)} & \colhead{(Jy)} &
\colhead{(Jy)}  & \colhead{(Jy)} & \colhead{(ergs s$^{-1}$)} &
\colhead{(Jy)} & \colhead{(Jy)} & \colhead{(mJy beam$^{-1}$)} &
\colhead{} \\
\colhead{(1)} & \colhead{(2)} & \colhead{(3)} & \colhead{(4)} &
\colhead{(5)} & \colhead{(6)} & \colhead{(7)} & \colhead{(8)} &
\colhead{(9)} & \colhead{(10)}  & \colhead{(11)}
}
\startdata
Mrk 335  & 0.025  & 0.30 & 0.38 & 0.34 & $<$0.57 & 44.2
(10.6) & 0.21 (i) & 0.15 & 1.89 & CfA, 12$\mu$m \\
NGC 863 (Mrk 590) & 0.027 &  0.19 & 0.22 & 0.49 & 1.46 & 44.2
(10.6) & 0.15 (ii) & \nodata & 3.44 & CfA \\
NGC 3786 (Mrk 744) & 0.009 &  0.10 \tablenotemark{a} &
0.39 \tablenotemark{a}& 1.20 \tablenotemark{a} & 3.00 \tablenotemark{a}
& 43.4 (9.8) & 0.059 (ii) & \nodata &  0.86 & CfA \\
NGC 4235 & 0.008 & $<$0.13 & $<$0.16 & 0.32 & 0.65 & 
42.5--43.0 (8.9--9.4) & 0.034 (ii) & \nodata  & 5.22 & CfA \\
NGC 4253 (Mrk 766) \tablenotemark{b} & 0.013 &  0.39 & 1.30 & 4.03 &
4.66 & 44.2 (10.6) & 0.24 (ii) & 0.25 & 4.77 & CfA, 12$\mu$m \\
NGC 5548 & 0.017 &  0.40 & 0.77 & 1.07 & 1.61 & 44.2 (10.6) & 0.20 (iii)
& 0.29 & 2.07 & CfA, 12$\mu$m  \\
Mrk 817  & 0.031 &  0.34 & 1.18 & 2.12 & 2.27 & 44.8 (11.2) & \nodata & 0.23 &
3.45 & CfA, 12$\mu$m \\
NGC 5940 & 0.034 &  $<$0.17 & 0.11 & 0.74 & 1.75 & 
44.3--44.4 (10.7--10.8) & 0.026 (ii) & \nodata & \nodata &  CfA \\
2237+07 (UGC 1238) & 0.025 &  0.20 & 0.34 & 0.83 & $<$2.49 & 
44.2--44.3 (10.6--10.7) & 0.063 (ii) & \nodata &  1.73 & CfA \\
NGC 7469 & 0.016 &  1.35 & 5.79 & 25.87 & 34.90 & 45.1 (11.5) & 0.60 (i)
& 0.31 &  13.63 & CfA, 12$\mu$m \\
Mrk 530 (NGC 7603) & 0.029 &  0.16 & $<$0.25 & 0.85 & 2.04 & 
44.3--44.4 (10.7--10.8) & 0.077 (i) & 0.10 &  2.78 & CfA, 12$\mu$m \\
\hline
NGC 931 (Mrk 1040) & 0.016 &  0.61 & 1.32 & 2.56 & 4.55 & 44.4 (10.8) 
& 0.33 (ii) & 0.21 & \nodata & 12$\mu$m \\
F03450+0055 & 0.031 &  0.28 & 0.51 & 0.47 & $<$3.24  & 
44.4--44.6 (10.8--11.0) & 0.17 (ii) & 0.10 & 4.2 & 12$\mu$m \\
3C 120   & 0.033 &  0.29 & 0.64 & 1.28 & 2.79 & 44.7 (11.1) & 0.22 (i) 
& 0.11 & \nodata & 12$\mu$m \\
Mrk 618  & 0.035 &  0.34 & 0.79 & 2.71 & 4.24 & 44.9 (11.3) & 0.27 (ii) &
$<$0.06 & 1.7 & 12$\mu$m \\
MCG-5-13-17 & 0.013 &  0.22 & 0.57 & 1.40 & 1.99 & 43.8 (10.2) & \nodata &
0.14 & 1.4 & 12$\mu$m \\
Mrk 79   & 0.022 &  0.31 & 0.76 & 1.50 & 2.36 & 44.4 (10.8) & 0.26 (ii)
& 0.24 & 0.8 & 12$\mu$m \\
NGC 2639 & 0.011 &  0.16 & 0.21 & 1.99 & 7.06 & 43.8 (10.2) & 0.008 (ii)
& $<$0.14 &  90.6 & 12$\mu$m \\
Mrk 704  & 0.030 &  0.35 & 0.53 & 0.36 & $<$0.77 & 44.4--44.5
(10.8--10.9) & 0.27 (ii) & 0.30 &  0.7 & 12$\mu$m \\
NGC 2992 & 0.008 &  0.59 & 1.37 & 6.87 & 14.4 & 44.0 (10.4) & 0.20 (iv)
& 0.16 & 3.9 & 12$\mu$m \\
Mrk 1239 \tablenotemark{b} & 0.019 &  0.65 & 1.14 & 1.34 & $<$2.42 &
44.4 (10.8) & 0.60 (ii) & 0.29 &  6.8 & 12$\mu$m \\
UGC 7064 & 0.025 &  0.17 & 0.38 & 2.75 & 5.57 & 44.6 (11.0) & \nodata &
\nodata & \nodata & 12$\mu$m \\
MCG-2-33-34 & 0.014 &  0.17 & 0.37 & 1.16 & 2.22 & 43.8 (10.2) & \nodata &
$<$0.03 & 0.6 & 12$\mu$m \\
IC 4329A & 0.016 &  1.08 & 2.21 & 2.03 & 1.66 & 44.5 (10.9) & 0.77 (i) &
0.92 & 4.9 & 12$\mu$m \\
Mrk 509 & 0.035 &  0.32 & 0.70 & 1.36 & 1.52 & 44.8 (11.2) & 0.24 (ii) &
0.20 & 1.8 & 12$\mu$m \\
\enddata

\tablecomments{
Column (1): Object. CfA Seyfert 1s are placed first. Ordered by right
ascension in each Seyfert 1 sample.
Column (2): Redshift.
Columns (3)--(6): f$_{12}$, f$_{25}$, f$_{60}$, and f$_{100}$ are
{\it IRAS FSC}
fluxes at 12 $\mu$m, 25 $\mu$m, 60 $\mu$m, and 100 $\mu$m, respectively.
Column (7): Logarithm of infrared (8$-$1000 $\mu$m) luminosity
in ergs s$^{-1}$ calculated with
$L_{\rm IR} = 2.1 \times 10^{39} \times$ D(Mpc)$^{2}$
$\times$ (13.48 $\times$ $f_{12}$ + 5.16 $\times$ $f_{25}$ +
$2.58 \times f_{60} + f_{100}$) ergs s$^{-1}$ \citep{san96}.
The values in parenthesis are logarithm of infrared luminosities in units of
solar luminosities. 
Column (8): Ground-based $N$-band (10.6 $\mu$m) photometric measurement 
using a single element detector ($>$4$''$ aperture) and references. (i):
Rieke 1978; (ii): Maiolino et al. 1995; (iii) Kleinmann \& Low 1970;
(iv) Glass, Moorwood, \& Eichendorf 1982.
Column (9): Ground-based $N$-band (10.6 $\mu$m) photometry measured
using a two-dimensional camera with a 1$\farcs$5 aperture \citep{gor04}.
Column (10): Peak radio intensity at 8.4 GHz, measured with 0$\farcs$25
resolution radio images \citep{kuk95,the00}.
Column (11): CfA \citep{huc92} or 12 $\mu$m \citep{rus93} Seyfert 1.
}

\tablenotetext{a}{ISO measurements \citep{per01}. The f$_{\rm 12}$ and
f$_{\rm 100}$ are extrapolated and interpolated, respectively, from
measurements at other wavelengths.}

\tablenotetext{b}{Observed by \citet{rod03}.}

\end{deluxetable}

\clearpage

\begin{deluxetable}{llcclcccc}
\tabletypesize{\scriptsize}
\tablecaption{Observing Log \label{tbl-2}}
\tablenum{2}
\tablewidth{0pt}
\tablehead{
\colhead{} & \colhead{Date} & \colhead{Telescope \&} &
\colhead{Integration} & \multicolumn{5}{c}{Standard Star} \\
\cline{5-9} \\
\colhead{Object} & \colhead{(UT)} & \colhead{Instrument} &
\colhead{Time (min)} & \colhead{Star Name} & \colhead{$L$-mag} &
\colhead{$K$-mag} & \colhead{Type} & \colhead{T$_{\rm eff}$ (K)} \\
\colhead{(1)} & \colhead{(2)} & \colhead{(3)} & \colhead{(4)}
& \colhead{(5)} & \colhead{(6)}  & \colhead{(7)} & \colhead{(8)} &
\colhead{(9)}
}
\startdata
Mrk 335  & 2003 Sep 6 & IRTF SpeX & 40 & HR 217 & 5.1 & 5.2 & F8V & 6000\\
NGC 863  & 2003 Sep 7 & IRTF SpeX & 60 & HR 650 & 4.1 & 4.2 & F8V & 6000\\
NGC 3786 & 2003 Mar 19 & IRTF SpeX & 30 & HR 4412 & 5.0 & 5.0 & F7V & 6240\\
NGC 4235 & 2003 Mar 21 & IRTF SpeX & 20 & HR 4708 & 5.0 & 5.0 & F8V & 6000\\
NGC 5548 & 2003 Mar 21 & IRTF SpeX & 20 & HR 5423 & 4.7 & 4.8 & G5V & 5700\\
Mrk 817  & 2003 Mar 21 & IRTF SpeX & 20 & HR 5581 & 4.3 & 4.3 & F7V & 6240\\
NGC 5940 & 2003 Mar 21 & IRTF SpeX & 20 & HR 5659 & 5.1 & 5.1 & G5V & 5700\\
2237+07  & 2003 Sep 6 & IRTF SpeX & 40 & HR 8514 & 4.9 & 5.0 & F6V & 6400\\
NGC 7469 & 2002 Aug 19 & Subaru IRCS & 10 & HR 8653 & 4.7 & \nodata & G8IV
& 5400
\\ Mrk 530  & 2003 Sep 6 & IRTF SpeX & 40 & HR 8772 & 5.3 & 5.3 & F8V & 6000\\
\hline
NGC 931  & 2003 Sep 6 & IRTF SpeX & 40 & HR 720 & 4.4 & 4.5 & G0V & 5930\\
F03450+0055 & 2003 Sep 6 & IRTF SpeX & 40 & HR 962 \tablenotemark{a} &
3.7 & 3.7 & F8V & 6000\\
3C 120   & 2003 Sep 7 & IRTF SpeX & 56 & HR 1232 \tablenotemark{b} & 3.9
& 4.0 & G9V & 5400\\
Mrk 618  & 2003 Sep 6 & IRTF SpeX & 60 & HR 1232 \tablenotemark{a}
& 3.9 & 4.0 & G9V & 5400\\
MCG-5-13-17 & 2003 Sep 7 & IRTF SpeX & 40 & HR 1785 & 5.2 & 5.3 & F6V & 6400\\
Mrk 79   & 2003 Mar 18 & IRTF SpeX & 30 & HR 3028 & 4.8 & 4.8 & F6V & 6400\\
NGC 2639 & 2003 Mar 21 & IRTF SpeX & 40 & HR 3309 & 4.7 & 4.7 & G5V & 5700\\
Mrk 704  & 2003 Mar 19 & IRTF SpeX & 30 & HR 3650 & 4.6 & 4.6 & G9V & 5400\\
NGC 2992 & 2003 Mar 19 & IRTF SpeX & 30 & HR 3901 & 5.0 & 5.1 & F8V & 6000\\
UGC 7064 & 2003 Mar 21 & IRTF SpeX & 30 & HR 4550 & 4.6 & 4.6 & G8V & 5400\\
MCG-2-33-34 & 2003 Mar 21 & IRTF SpeX & 20 & HR 4856 & 4.9 & 4.9 & F7V
& 6240\\
IC 4329A & 2003 Mar 20 & IRTF SpeX & 12 & HR 5212 \tablenotemark{c} & 4.8
& 4.8 & F7V &6240\\
Mrk 509 & 2003 Sep 6 & IRTF SpeX & 40 & HR 7715 \tablenotemark{d} & 4.5 &
4.5 & F7V & 6240 \\
\enddata

\tablecomments{Column (1): Object.
Column (2): Observing date in UT.
Column (3): Instrument used for spectroscopy.
Column (4): Net on-source integration time in min.
Column (5): Standard star name.
Column (6): Adopted $L$-band magnitude.
Column (7): Adopted $K$-band magnitude.
Column (8): Stellar spectral type.
Column (9): Effective temperature.
}

\tablenotetext{a}{HR 720 is used as a standard star for the $K$-band
spectrum, because signals of HR 962 in the $K$-band are above the
linearity level of the SpeX array.}

\tablenotetext{b}{HR 1785 is used as a standard star for the $K$-band
spectrum.}

\tablenotetext{c}{Flux calibration was made using HR 4412.}

\tablenotetext{d}{HR 8772 is used as a standard star for the $K$-band
spectrum.}

\end{deluxetable}

\clearpage

\begin{deluxetable}{lccccc}
\tabletypesize{\scriptsize}
\rotate
\tablecaption{Properties of the Nuclear 3.3 $\mu$m PAH Emission Feature
\label{tbl-3}} 
\tablenum{3}
\tablewidth{0pt}
\tablehead{
\colhead{} & \colhead{f$_{\rm nuclear-3.3 PAH}$} &
\colhead{L$_{\rm nuclear-3.3 PAH}$} & \colhead{rest EW$_{3.3 \rm PAH}$} &
\colhead{L$_{\rm nuclear-3.3 PAH}$/L$_{\rm IR}$}  &
\colhead{S$_{\rm nuclear-3.3 PAH}$}  \\ 
\colhead{Object} & \colhead{($\times$ 10$^{-14}$ ergs s$^{-1}$ cm$^{-2}$)} &
\colhead{($\times$ 10$^{39}$ ergs s$^{-1}$)} & \colhead{(nm)} &
\colhead{($\times$ 10$^{-3}$)} &
  \colhead{($\times$ 10$^{39}$ ergs s$^{-1}$ kpc$^{-2}$)} \\
\colhead{(1)} & \colhead{(2)} & \colhead{(3)} & \colhead{(4)}
& \colhead{(5)} & \colhead{(6)}
}
\startdata
Mrk 335  & 2.5    & 31     & 1.4    & 0.19       & 36 \\
NGC 863  & $<$2.3 & $<$33  & $<$5.4 & $<$0.20    & $<$33 \\
NGC 3786 & $<$6.4 & $<$10  & $<$7.0 & $<$0.38    & $<$91 \\
NGC 4235 & 3.4    & 4.4    & 8.7    & 0.5--1.3   & 49 \\
NGC 4253 & 6.5 \tablenotemark{a} & 20.7 \tablenotemark{a}
& 40 \tablenotemark{a} & 0.14 \tablenotemark{a}  & 230 \\
NGC 5548 & $<$4.3 & $<$25 & $<$4.3 & $<$0.18    & $<$61 \\
Mrk 817  & 5.4    & 100    & 4.3    & 0.16       & 78 \\
NGC 5940 & 2.7    &  63    & 9.8    & 0.23--0.35 & 41 \\
2237+07  & 4.4    &  54    & 9.8    & 0.27--0.36 & 64 \\
NGC 7469 & $<$7.0 & $<$35  & $<$2.4 & 0.026      & $<$390 \\
Mrk 530  & 4.6    &  76    & 1.5    & 0.33--0.39 & 67 \\
\hline
NGC 931  & $<$4.3 & $<$22  & $<$2.3 & $<$0.091   & $<$61 \\
F03450+0055 & $<$3.8 & $<$73 & $<$2.2 & $<$0.28  & $<$57 \\
3C 120   & 4.9    & 110    & 1.9    & 0.21       & 76 \\
Mrk 618  & 2.9    & 71     & 3.1    & 0.081      & 44 \\
MCG-5-13-17 & $<$4.8 & $<$16 & $<$9.5 & $<$0.23  & $<$67 \\
Mrk 79   & $<$4.9 & $<$47  & $<$2.6 & $<$0.19    & $<$71 \\
NGC 2639 & $<$1.7 & $<$3.9 & $<$9.5 & $<$0.06    & $<$23 \\
Mrk 704  & $<$2.2 & $<$39  & $<$0.9 & $<$0.15    & $<$32 \\
NGC 2992 & $<$9.1 & $<$12  & $<$2.7 & $<$0.11    & $<$133 \\
Mrk 1239 & $<$1.2 \tablenotemark{a} & $<$8.9 \tablenotemark{a} &
$<$2.6 \tablenotemark{a} & $<$0.04 \tablenotemark{a}  & $<$47 \\
UGC 7064 & $<$2.4 & $<$30  & $<$9.2 & $<$0.079   & $<$35 \\
MCG-2-33-34 & 3.8 & 15     & 10.3   & 0.22       & 54 \\
IC 4329A & $<$16  & $<$78  & $<$1.3 & $<$0.27    & $<$220 \\
Mrk 509  & 4.3    & 100    & 1.6    & 0.18       & 62 \\
\enddata

\tablecomments{Column (1): Object.
Column (2): Observed nuclear 3.3 $\mu$m PAH flux.
Column (3): Observed nuclear 3.3 $\mu$m PAH luminosity.
Column (4): Rest frame equivalent width of the 3.3 $\mu$m PAH emission.
Column (5): Observed nuclear 3.3 $\mu$m PAH to infrared luminosity ratio
             in units of 10$^{-3}$, a typical value for starburst
             galaxies.
Column (6): Surface brightness of the nuclear 3.3 $\mu$m PAH emission.
}

\tablenotetext{a}{Taken from \citet{rod03}.}

\end{deluxetable}

\clearpage

\begin{deluxetable}{lc}
\tabletypesize{\scriptsize}
\tablecaption{Nuclear $K - L$ Color \label{tbl-4}}
\tablenum{4}
\tablewidth{0pt}
\tablehead{
\colhead{Object} & \colhead{$K - L$ [mag]} \\
\colhead{(1)} & \colhead{(2)} 
}
\startdata
Mrk 335  & 1.9 \\
NGC 863  & 0.6 \\
NGC 3786 & 1.3 \\
NGC 4235 & 1.1 \\
NGC 5548 & 1.7 \\
Mrk 817  & 1.7 \\
NGC 5940 & 1.3 \\
2237+07  & 1.2 \\
Mrk 530  & 1.7 \\
NGC 931  & 1.7 \\
F03450+0055  \tablenotemark{a} & 1.8 \\
3C 120  \tablenotemark{a} & 2.0 \\
Mrk 618 \tablenotemark{a} & 1.6 \\
MCG-5-13-17 & 1.0 \\
Mrk 79   & 1.7 \\
NGC 2639 & 0.3 \\
Mrk 704 & 1.7 \\
NGC 2992 & 1.6 \\
UGC 7064 & 1.1 \\
MCG-2-33-34 & 1.1 \\
IC 4329A & 2.1 \\
Mrk 509 \tablenotemark{a} & 1.8 \\
\enddata

\tablecomments{Column (1): Object.
Column (2): Nuclear $K - L$ magnitude measured with our slit
spectroscopy.
}

\tablenotetext{a}{$K$- and $L$-band magnitudes were measured using
different standard stars.}

\end{deluxetable}

\clearpage

\begin{deluxetable}{lcccc}
\tabletypesize{\scriptsize}
\tablecaption{Br$\gamma$ Emission Line \label{tbl-5}}
\tablenum{5}
\tablewidth{0pt}
\tablehead{
\colhead{} & \colhead{FWHM} & \colhead{f$_{\rm Br \gamma}$} &
\colhead{L$_{\rm Br \gamma}$} &
\colhead{rest EW$_{\rm Br \gamma}$} \\
\colhead{Object} & \colhead{(km s$^{-1}$)}
& \colhead{($\times$ 10$^{-14}$ ergs s$^{-1}$ cm$^{-2}$)} &
\colhead{($\times$ 10$^{40}$ ergs s$^{-1}$)} & \colhead{(\AA)} \\
\colhead{(1)} & \colhead{(2)} & \colhead{(3)} & \colhead{(4)} & \colhead{(5)}
}
\startdata
Mrk 335  & 1230 & 1.9  & 2.3   & 9.0 \\
2237+07  & 2250 & 0.9  & 1.1   & 10.3 \\
Mrk 530  &  920 & 0.4  & 0.7   & 1.1 \\
NGC 931  & 2390 & 2.2  & 1.1   & 9.3 \\
F03450   & 1200 & 1.8  & 3.4   & 8.9 \\
3C 120   & 4970 & 4.8  & 10.4  & 18.2 \\
Mrk 618  & 2200 & 1.4  & 3.5   & 10.7 \\
Mrk 79   & 2590 & 2.1  & 2.0   & 8.5  \\
NGC 2992 & 2000 & 4.5  &  0.6  & 9.7 \\
MCG-2-33-34 & 2240 & 1.9 & 0.7 & 25.1 \\
IC 4329A & 4660 & 21.5 & 10.7  & 20.2 \\
Mrk 509  & 3020 & 5.3  & 13.0  & 16.6 \\
\enddata

\tablecomments{Column (1): Object.
Column (2): Full-width at half maximum of the Br$\gamma$ emission line.
Column (3): Observed Br$\gamma$ emission flux.
Column (4): Observed Br$\gamma$ emission luminosity.
Column (5): Rest frame equivalent width of the Br$\gamma$ emission.
}

\end{deluxetable}

\clearpage

\begin{deluxetable}{lc}
\tabletypesize{\scriptsize}
\tablecaption{CO Absorption Feature \label{tbl-6}}
\tablenum{6}
\tablewidth{0pt}
\tablehead{
\colhead{Object} & \colhead{CO$_{\rm spec}$} \\
\colhead{(1)} & \colhead{(2)}
}
\startdata
NGC 863  & 0.04 \\
NGC 3786 & 0.10 \\
NGC 4235 & 0.08 \\
2237+07  & 0.05 \\
MCG-5-13-17 & 0.10 \\
NGC 2639 & 0.21 \\
UGC 7064 & 0.10 \\
MCG-2-33-34 & 0.10 \\
\enddata

\tablecomments{Column (1): Object.
Column (2): Observed spectroscopic CO index.
}

\end{deluxetable}

\clearpage

\begin{deluxetable}{lccccc}
\tabletypesize{\scriptsize}
\tablecaption{Detailed Information on Seyfert 2 Galaxies for Comparison
\label{tbl-7}}
\tablenum{7}
\tablewidth{0pt}
\tablehead{
\colhead{Object} & \colhead{Redshift} & 
\colhead{L$_{\rm nuclear-3.3 PAH}$} &
\colhead{f$_{\rm N}$ (1$\farcs$5)}  &
\colhead{f$_{8.4 \rm GHz}$} & \colhead{Remarks} \\
\colhead{(Sy2)} & \colhead{} &
\colhead{($\times$ 10$^{39}$ ergs s$^{-1}$)} &
\colhead{(Jy)} & \colhead{(mJy beam$^{-1}$)} & \colhead{} \\
\colhead{(1)} & \colhead{(2)} & \colhead{(3)} & \colhead{(4)} &
\colhead{(5)} & \colhead{(6)}
}
\startdata
Mrk 334  & 0.022 & 58     & \nodata & 1.33 & CfA \\
Mrk 993  & 0.015 & $<$9.4 & \nodata & 0.44 & CfA \\
Mrk 573  & 0.017 & $<$18  & \nodata & 0.61 & CfA \\
NGC 1144 & 0.029 & $<$12  & \nodata & 2.03 & CfA, 12$\mu$m \\
NGC 4388 & 0.008 & $<$4.7 & 0.22 & 2.74 & CfA, 12$\mu$m \\
NGC 5252 & 0.023 & $<$56  & \nodata & 4.57 & CfA \\
NGC 5256 (Mrk 266SW) & 0.028 & 133 & 0.07 & 3.49 & CfA, 12$\mu$m \\
NGC 5347 & 0.008 & $<$7.9 & 0.21 & 0.7 & CfA, 12$\mu$m \\
NGC 5674 & 0.025 & $<$19  & \nodata & \nodata & CfA \\
NGC 5695 & 0.014 & $<$6.1 & \nodata & \nodata & CfA \\
NGC 5929 & 0.008 & $<$1.7 & $<$0.03 & 8.22 & CfA, 12$\mu$m \\
NGC 7674 & 0.029 & 136    & 0.25    & \nodata & CfA, 12$\mu$m \\
NGC 7682 & 0.017 & 11     & \nodata & 12.22 & CfA \\  \hline
Mrk 938  & 0.019 & 289    & 0.16 & 6.2 & 12$\mu$m \\
NGC 262 (Mrk 348) & 0.015 & $<$38 & 0.12 & 310.3 & 12$\mu$m \\
NGC 513  & 0.020 & $<$22  & $<$0.10 & 0.8 & 12$\mu$m \\
F01475$-$0740 & 0.017 & 22 & \nodata & 132.0 & 12$\mu$m \\
NGC 1125 & 0.011 & 24     & $<$0.06 & 4.2 & 12$\mu$m \\
NGC 1194 & 0.013 & $<$8.8 & \nodata & 0.7 & 12$\mu$m \\
NGC 1241 & 0.014 & $<$4.1 & $<$0.10 & 5.3 & 12$\mu$m \\
NGC 1320 (Mrk 607) & 0.010 & $<$12 & 0.23 & 0.7 & 12$\mu$m \\
F04385$-$0828 & 0.015 & 18 & 0.16 & 4.7 & 12$\mu$m \\
NGC 1667 & 0.015 & $<$7.1 & $<$0.10 & 0.4 & 12$\mu$m \\
NGC 3660 & 0.012 & 11     & $<$0.11 & \nodata & 12$\mu$m \\
NGC 4501 & 0.008 & $<$4.7 & $<$0.11 & \nodata & 12$\mu$m, CfA
\tablenotemark{a} \\
NGC 4968 & 0.010 & 17     & 0.25    & 2.8 & 12$\mu$m \\
MCG-3-34-64 & 0.017 & $<$18 & 0.59  & 24.5 & 12$\mu$m  \\
NGC 5135 & 0.014 & 58     & 0.12    & \nodata & 12$\mu$m \\
MCG-2-40-4 (NGC 5995) & 0.024 & $<$80 & 0.22   & 1.6 & 12$\mu$m \\
F15480$-$0344 & 0.030 & $<$59 & 0.10 & 9.5 & 12$\mu$m \\
NGC 7172 & 0.009 & $<$8.4 & 0.11    & 2.0 & 12$\mu$m \\
MCG-3-58-7 & 0.032 & $<$78 & 0.20   & 0.3 & 12$\mu$m \\
\enddata

\tablecomments{Column (1): Object name of a Seyfert 2 galaxy.
Column (2): Redshift.
Column (3): Observed nuclear 3.3 $\mu$m PAH emission luminosity, taken
from \citet{ima03}. 
Column (4): Observed nuclear $N$-band flux measured using a
two-dimensional camera with a 1$\farcs$5 aperture \citep{gor04}.
Column (5): Peak radio intensity at 8.4 GHz measured with 0$\farcs$25
spatial resolution \citep{kuk95,the00}.
Column (6): CfA or 12 $\mu$m Seyfert 2.
}

\tablenotetext{a}{NGC 4501 is classified as a LINER in the CfA sample
\citep{huc92}.}

\end{deluxetable}

\clearpage

\begin{figure}
\includegraphics[angle=-90,scale=.35]{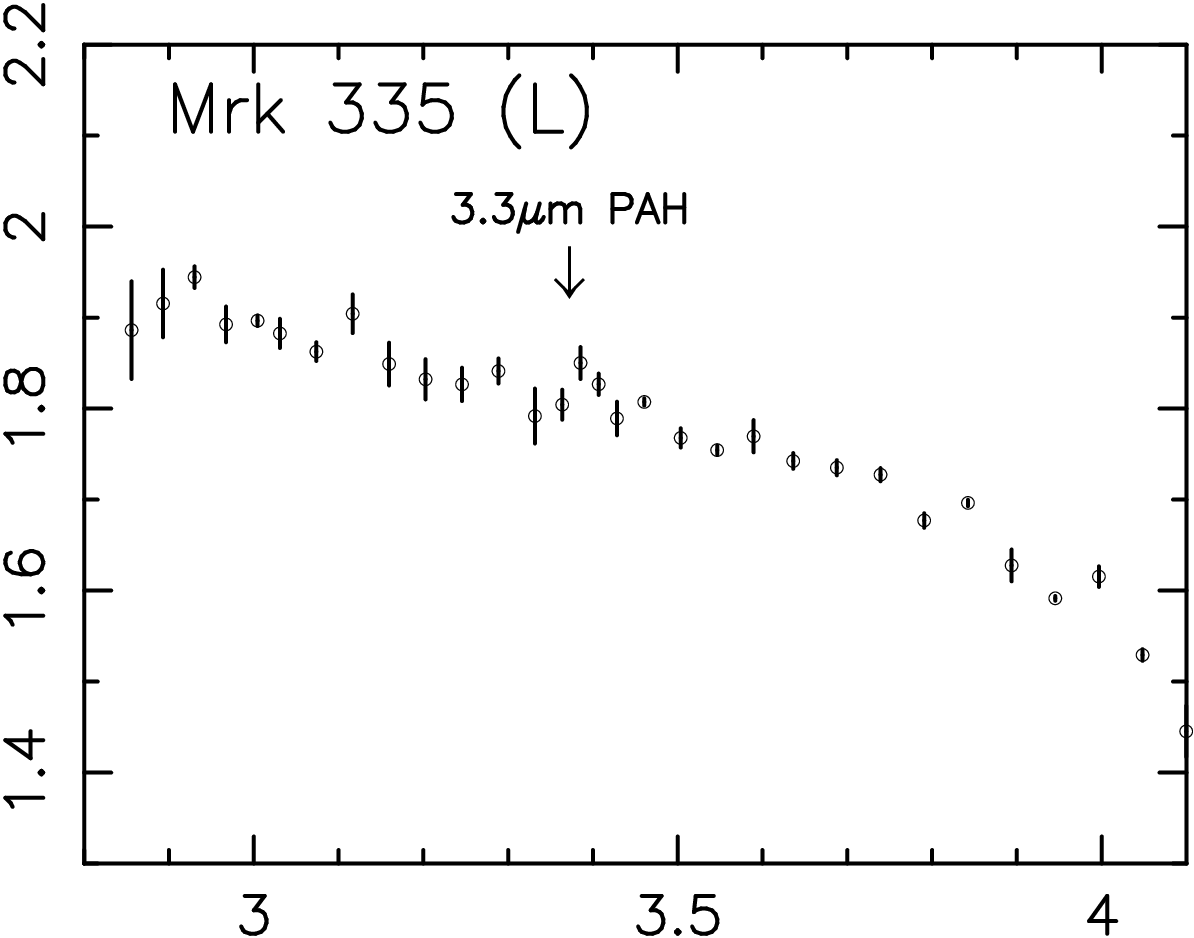} \hspace{0.3cm}
\includegraphics[angle=-90,scale=.35]{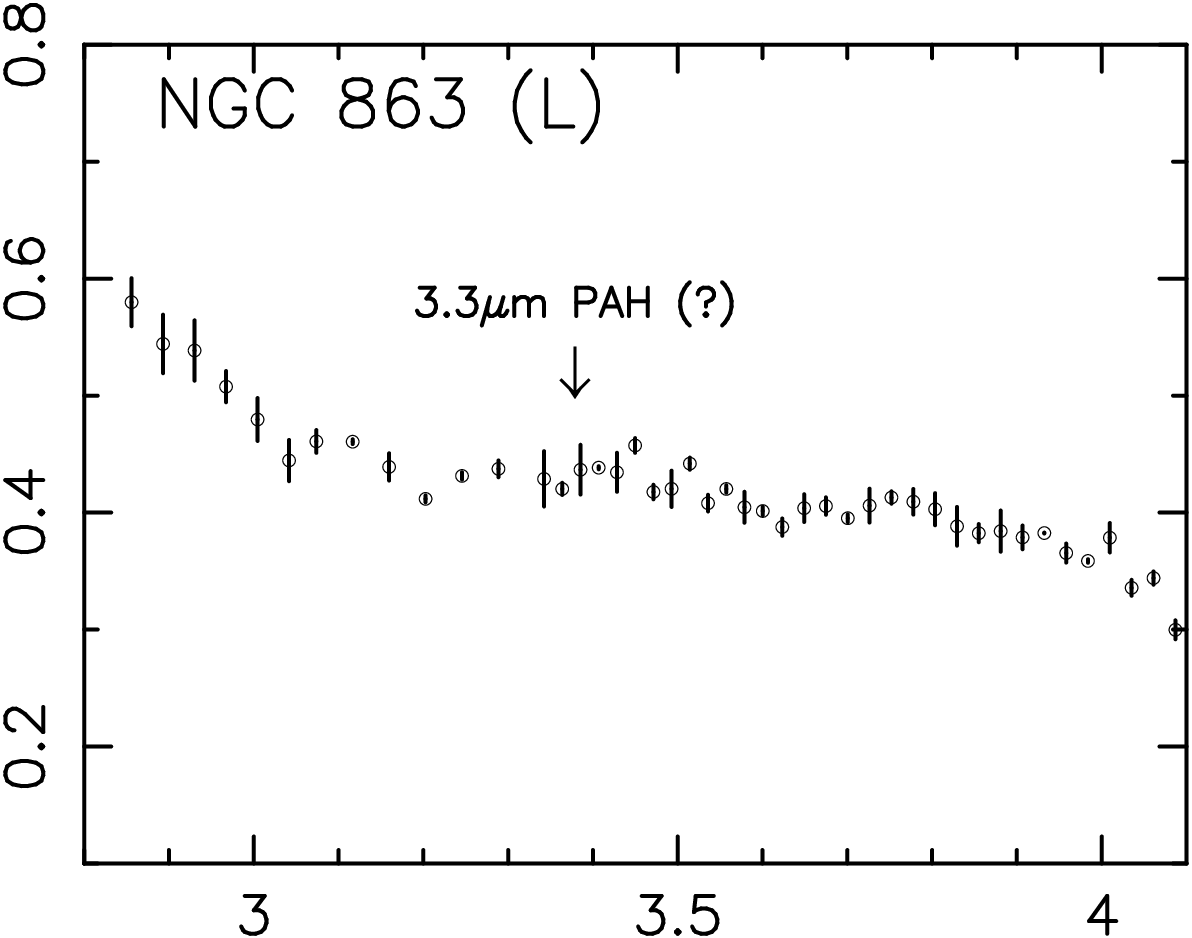} \\
\includegraphics[angle=-90,scale=.35]{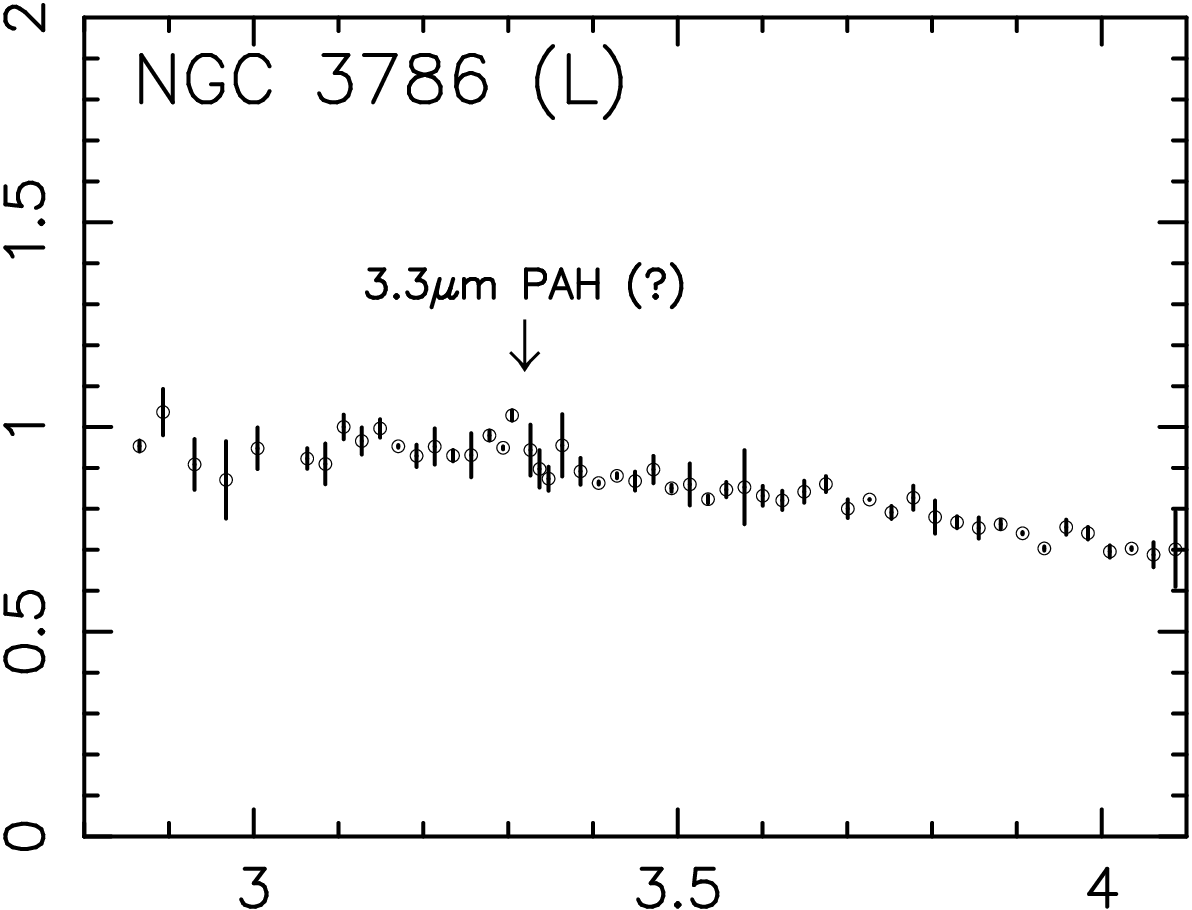} \hspace{0.3cm}
\includegraphics[angle=-90,scale=.35]{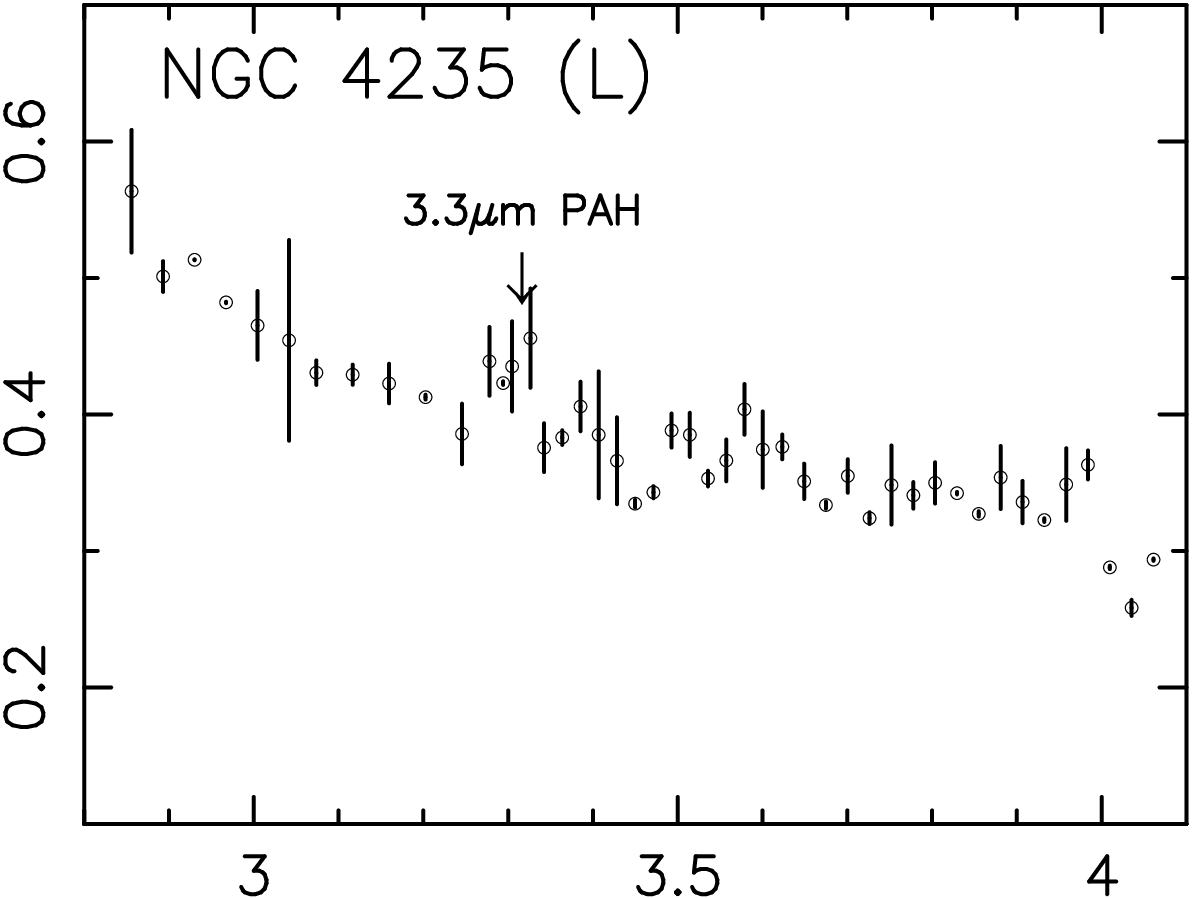} \\
\includegraphics[angle=-90,scale=.35]{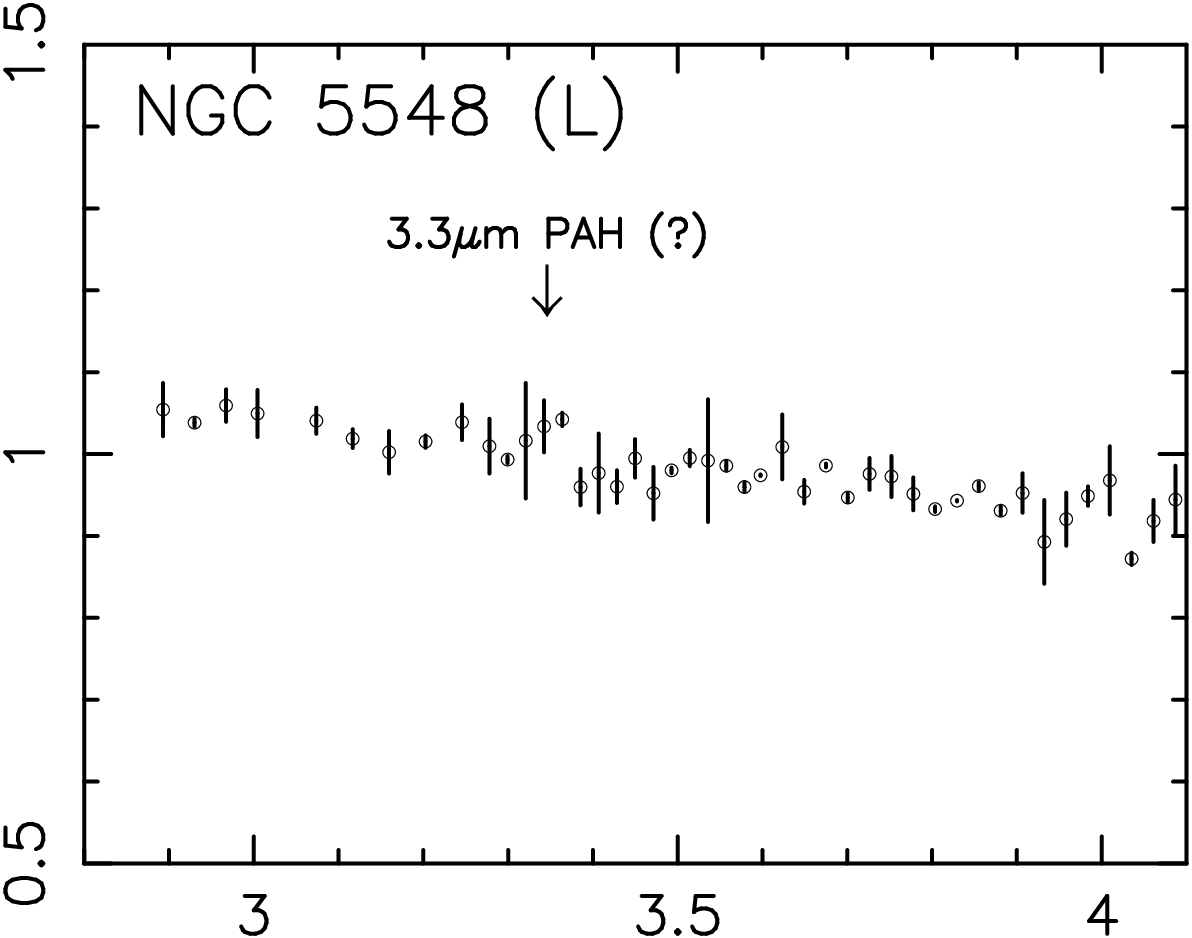} \hspace{0.3cm}
\includegraphics[angle=-90,scale=.35]{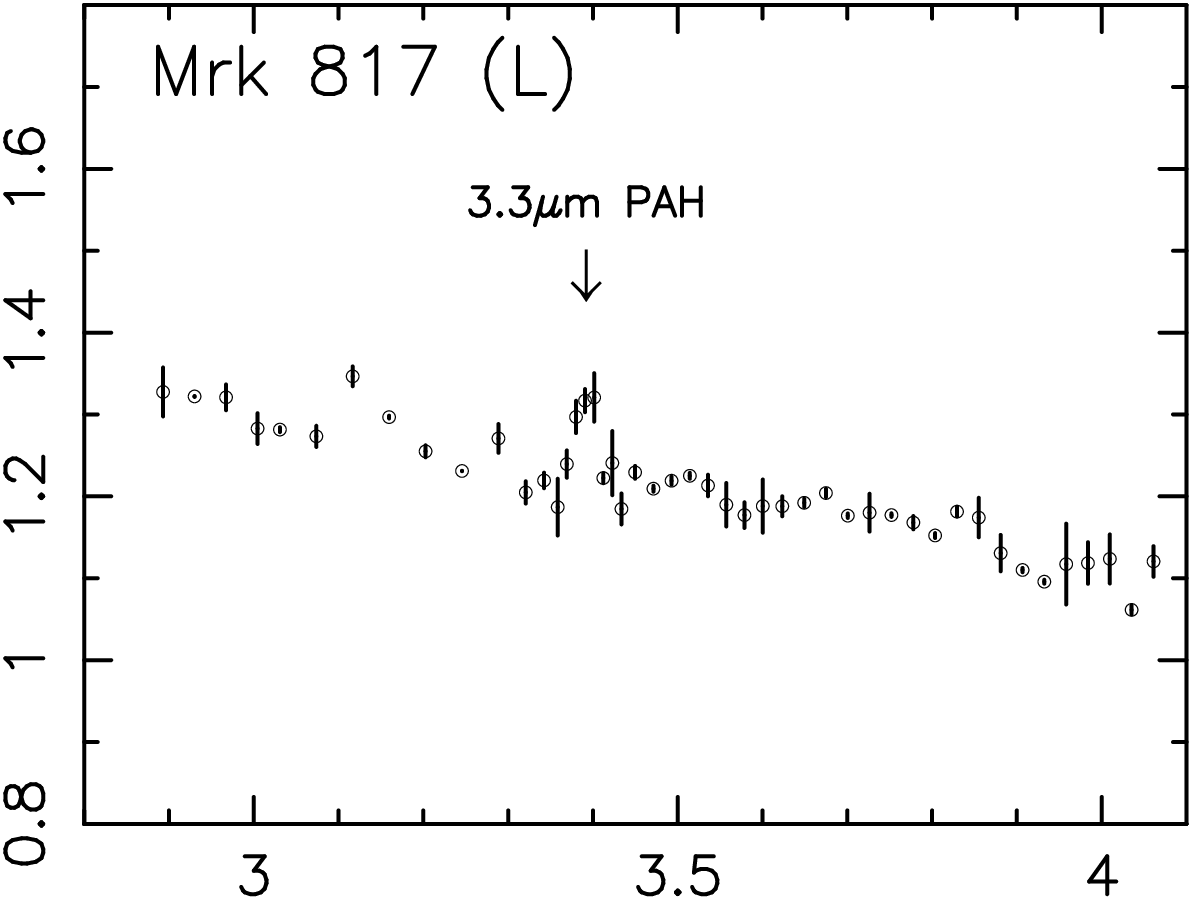} \\
\end{figure}
\clearpage
\begin{figure}
\includegraphics[angle=-90,scale=.35]{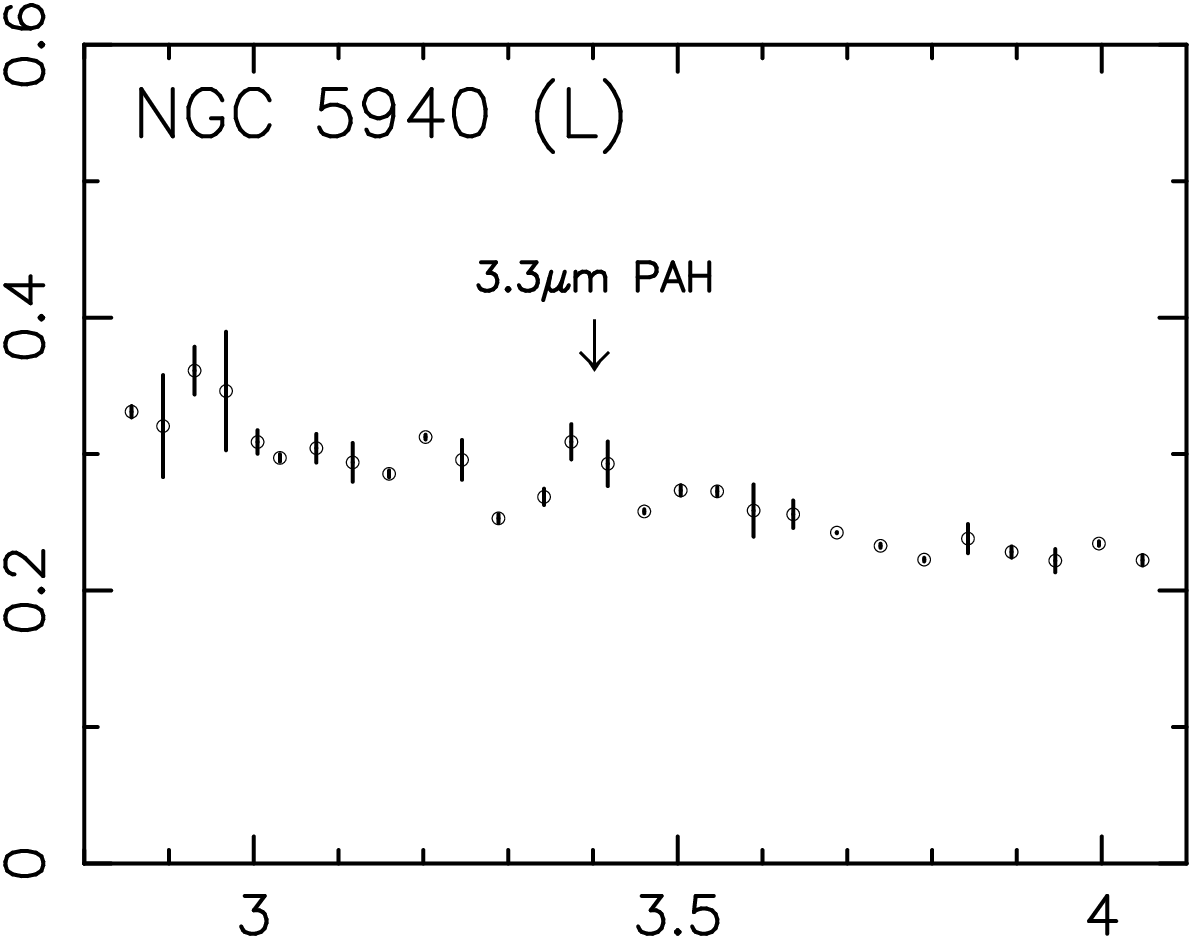} \hspace{0.3cm}
\includegraphics[angle=-90,scale=.35]{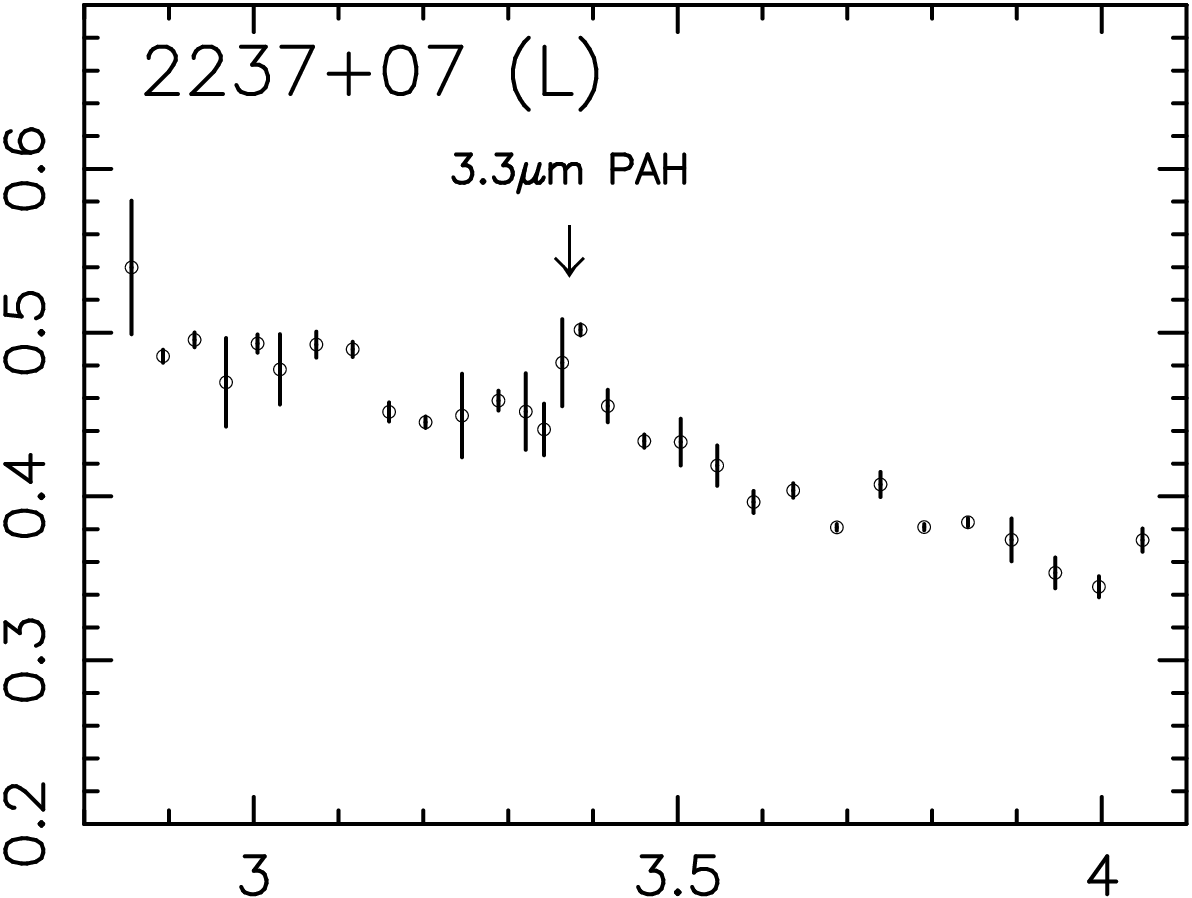} \\
\includegraphics[angle=-90,scale=.35]{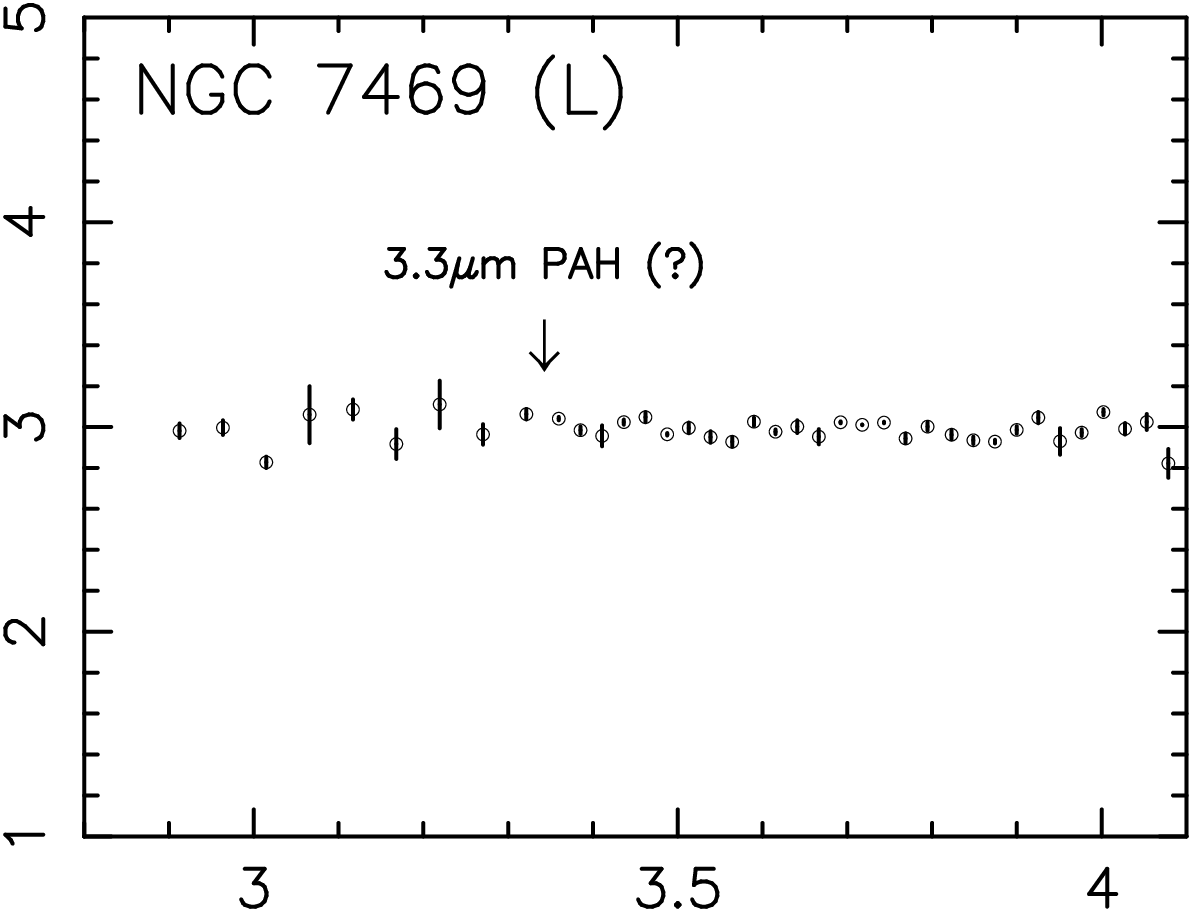} \hspace{0.3cm}
\includegraphics[angle=-90,scale=.35]{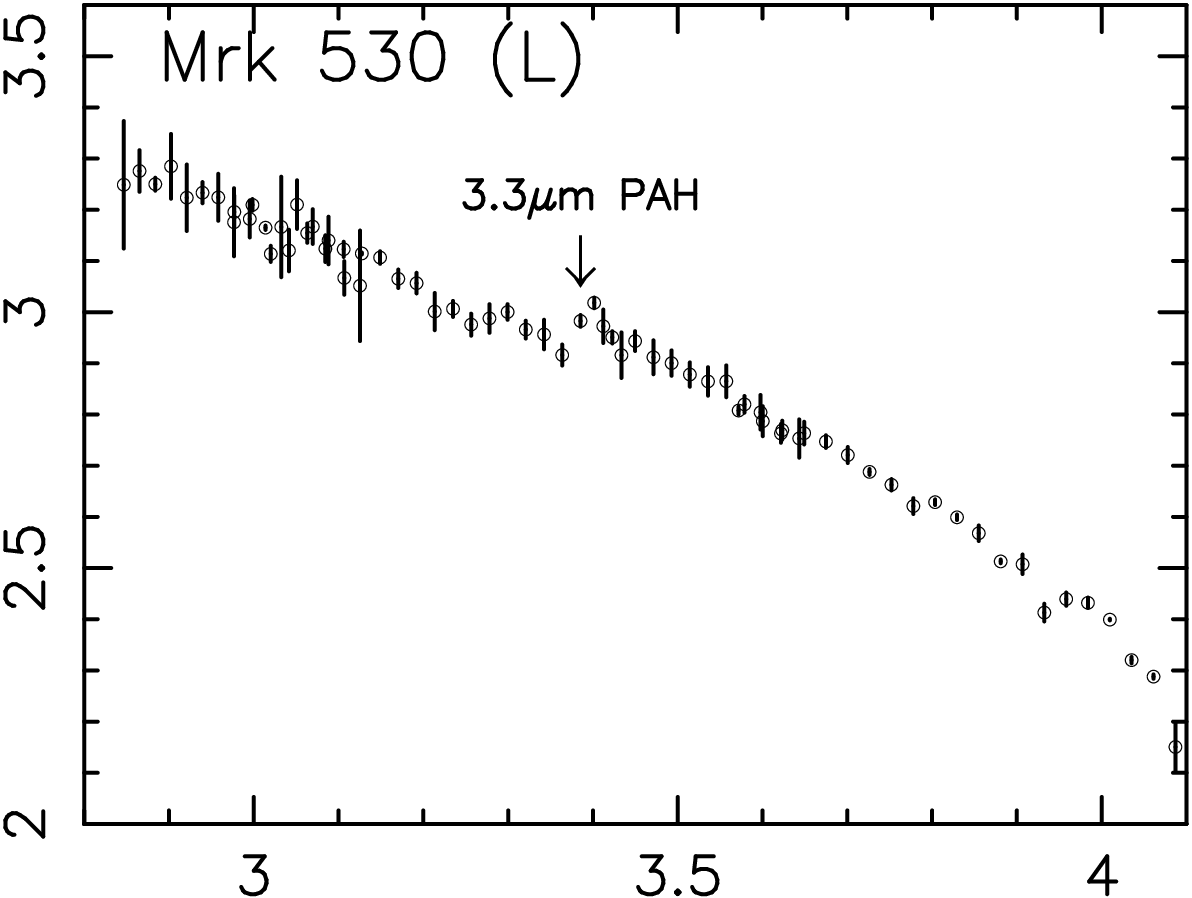} \\
\includegraphics[angle=-90,scale=.35]{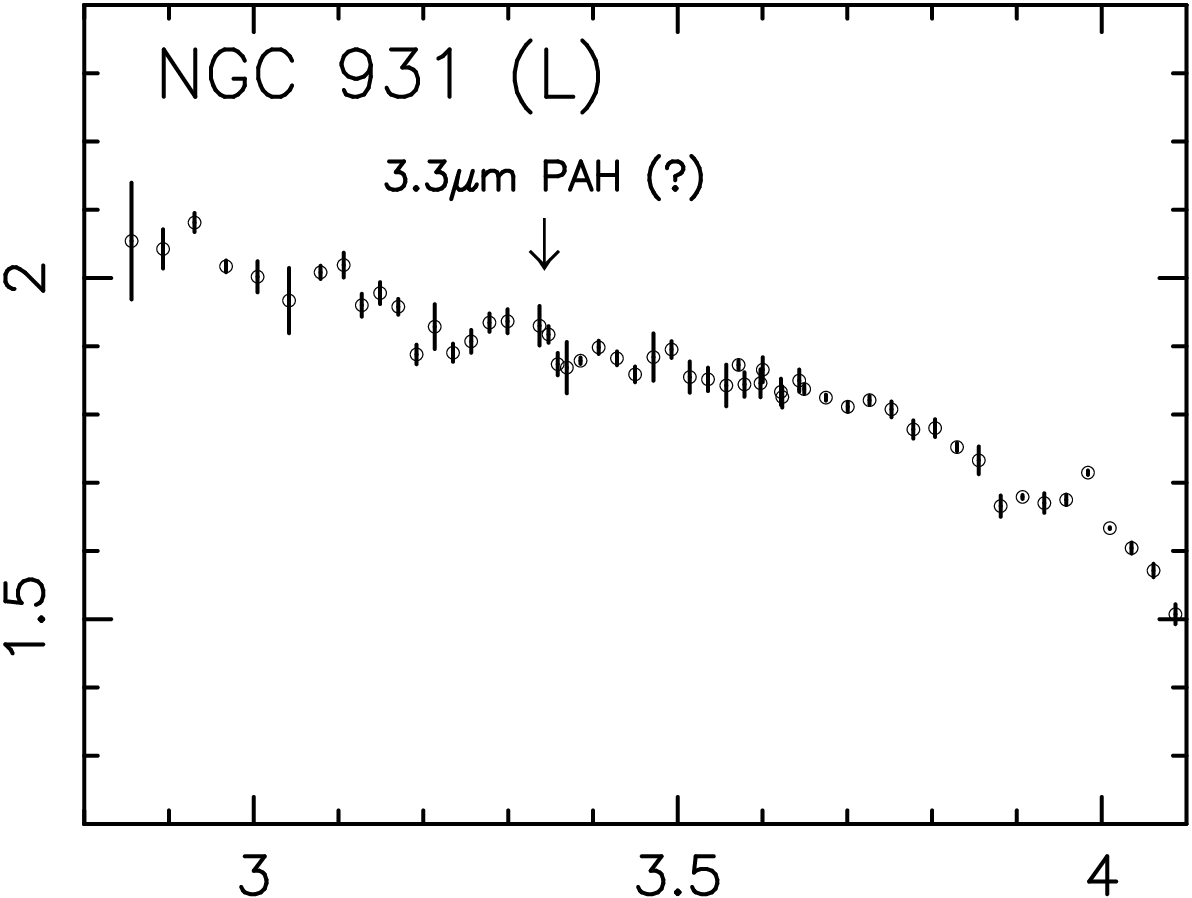} \hspace{0.3cm}
\includegraphics[angle=-90,scale=.35]{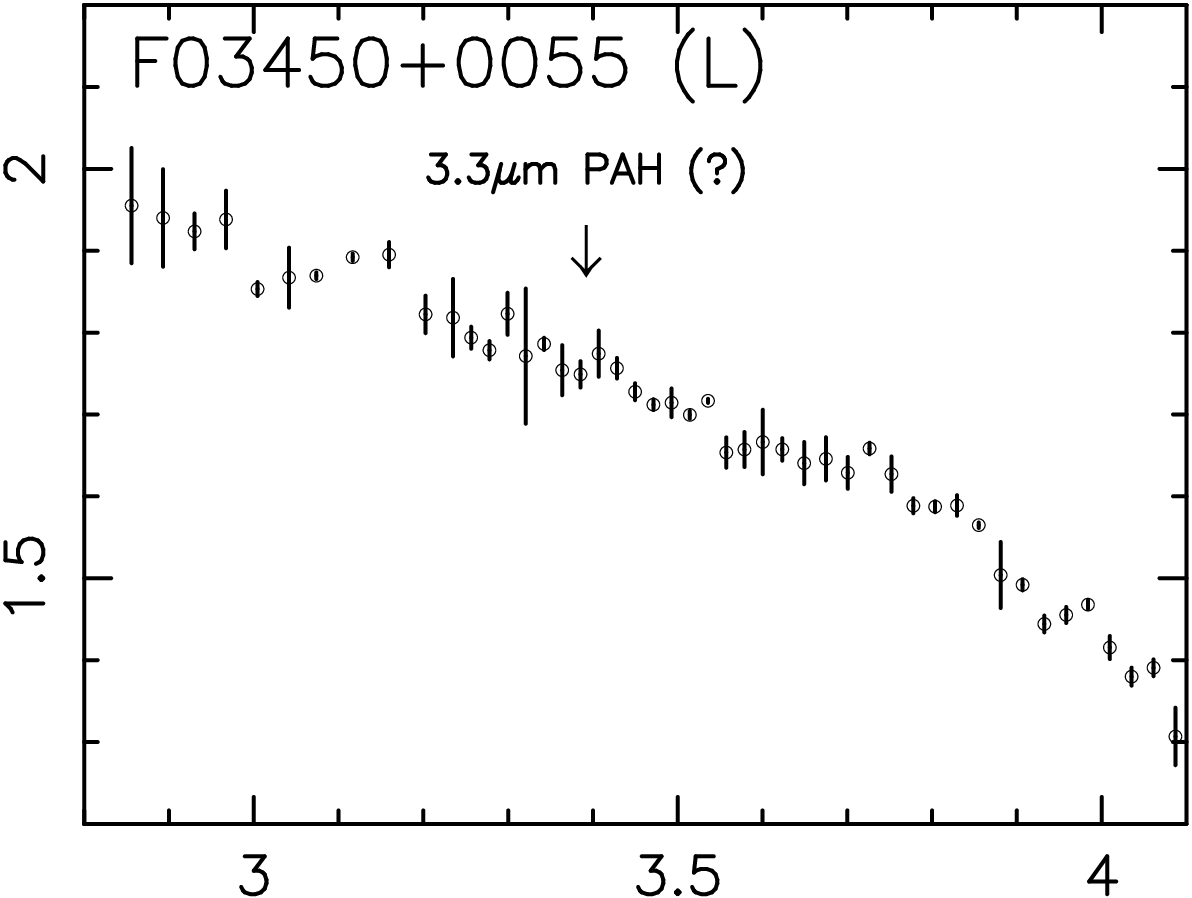} \\
\end{figure}
\clearpage
\begin{figure}
\includegraphics[angle=-90,scale=.35]{f1m.eps} \hspace{0.3cm}
\includegraphics[angle=-90,scale=.35]{f1n.eps} \\
\includegraphics[angle=-90,scale=.35]{f1o.eps} \hspace{0.3cm}
\includegraphics[angle=-90,scale=.35]{f1p.eps} \\
\includegraphics[angle=-90,scale=.35]{f1q.eps} \hspace{0.3cm}
\includegraphics[angle=-90,scale=.35]{f1r.eps} \\
\end{figure}
\clearpage
\begin{figure}
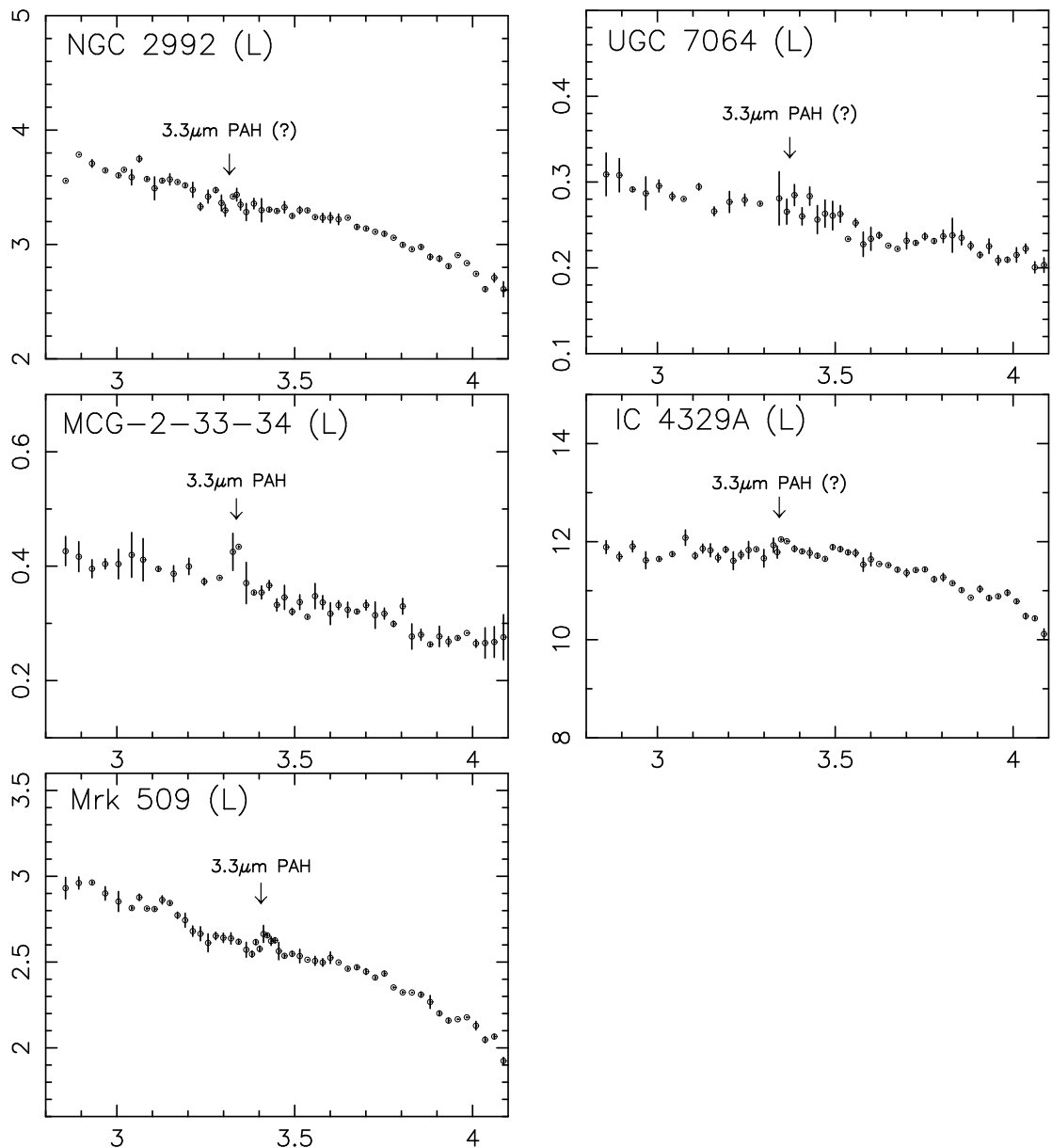

\includegraphics[angle=-90,scale=.35]{f1s.eps} \hspace{0.3cm}
\includegraphics[angle=-90,scale=.35]{f1t.eps} \\
\includegraphics[angle=-90,scale=.35]{f1u.eps} \hspace{0.3cm}
\includegraphics[angle=-90,scale=.35]{f1v.eps} \\
\includegraphics[angle=-90,scale=.35]{f1w.eps}
\caption{Infrared $L$-band (2.8--4.1 $\mu$m) slit spectra of the 23
Seyfert 1 nuclei.
The abscissa and ordinate are the observed wavelength in $\mu$m and
F$_{\lambda}$ in 10$^{-14}$ W m$^{-2}$ $\mu$m$^{-1}$, respectively.
Sources marked with ``3.3 $\mu$m PAH'' [``3.3 $\mu$m PAH (?)''] show
detectable [undetectable] 3.3 $\mu$m PAH emission features.
}
\end{figure}

\clearpage

\begin{figure}
\includegraphics[angle=-90,scale=.35]{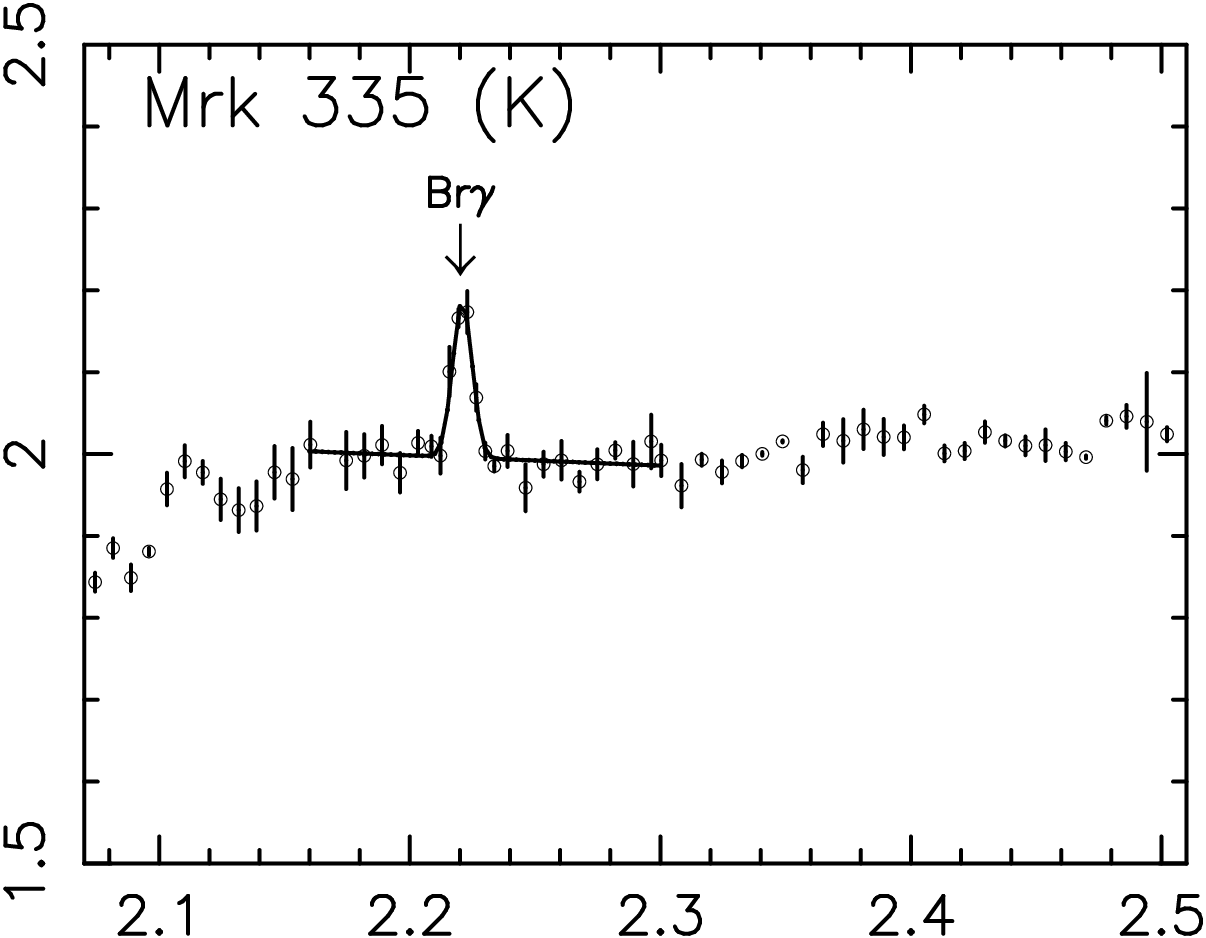} \hspace{0.3cm}
\includegraphics[angle=-90,scale=.35]{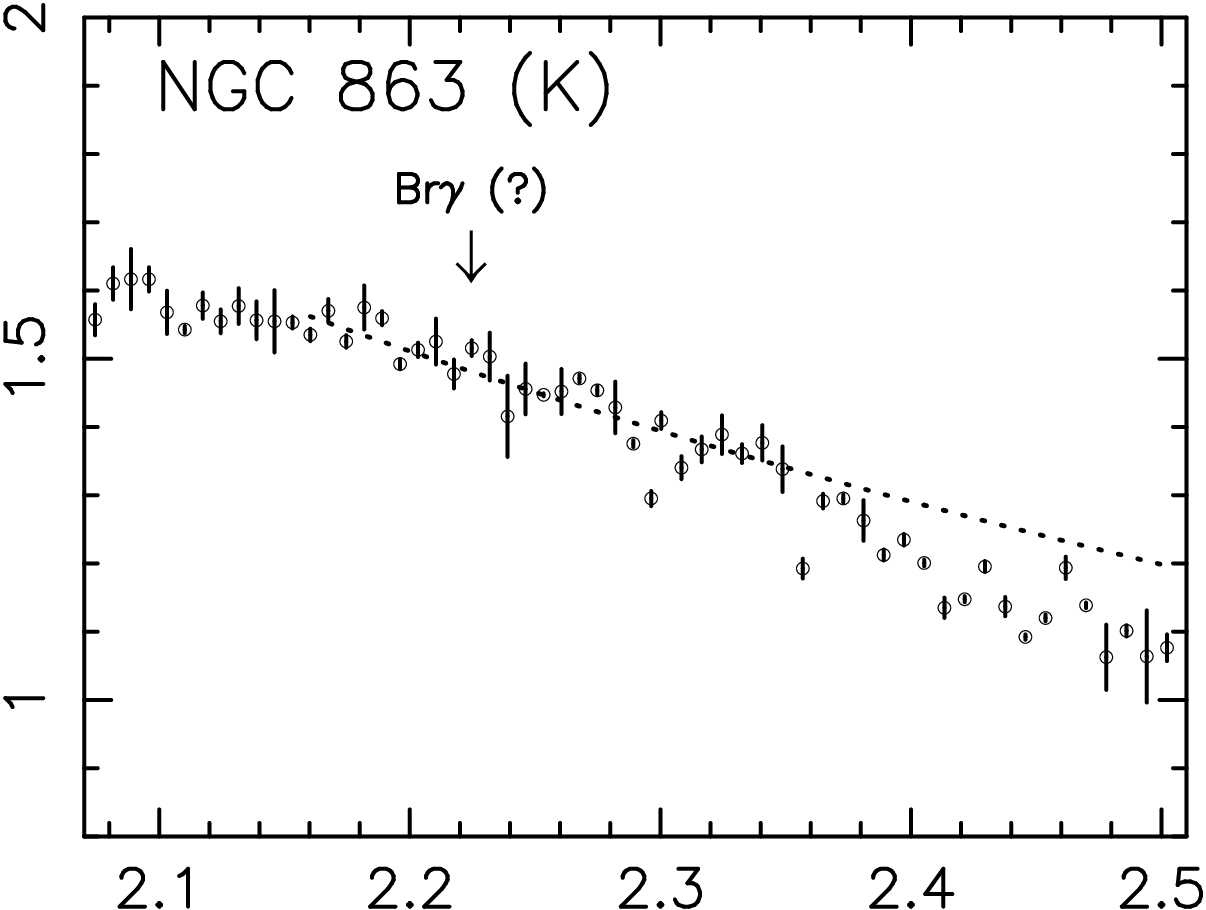} \\
\includegraphics[angle=-90,scale=.35]{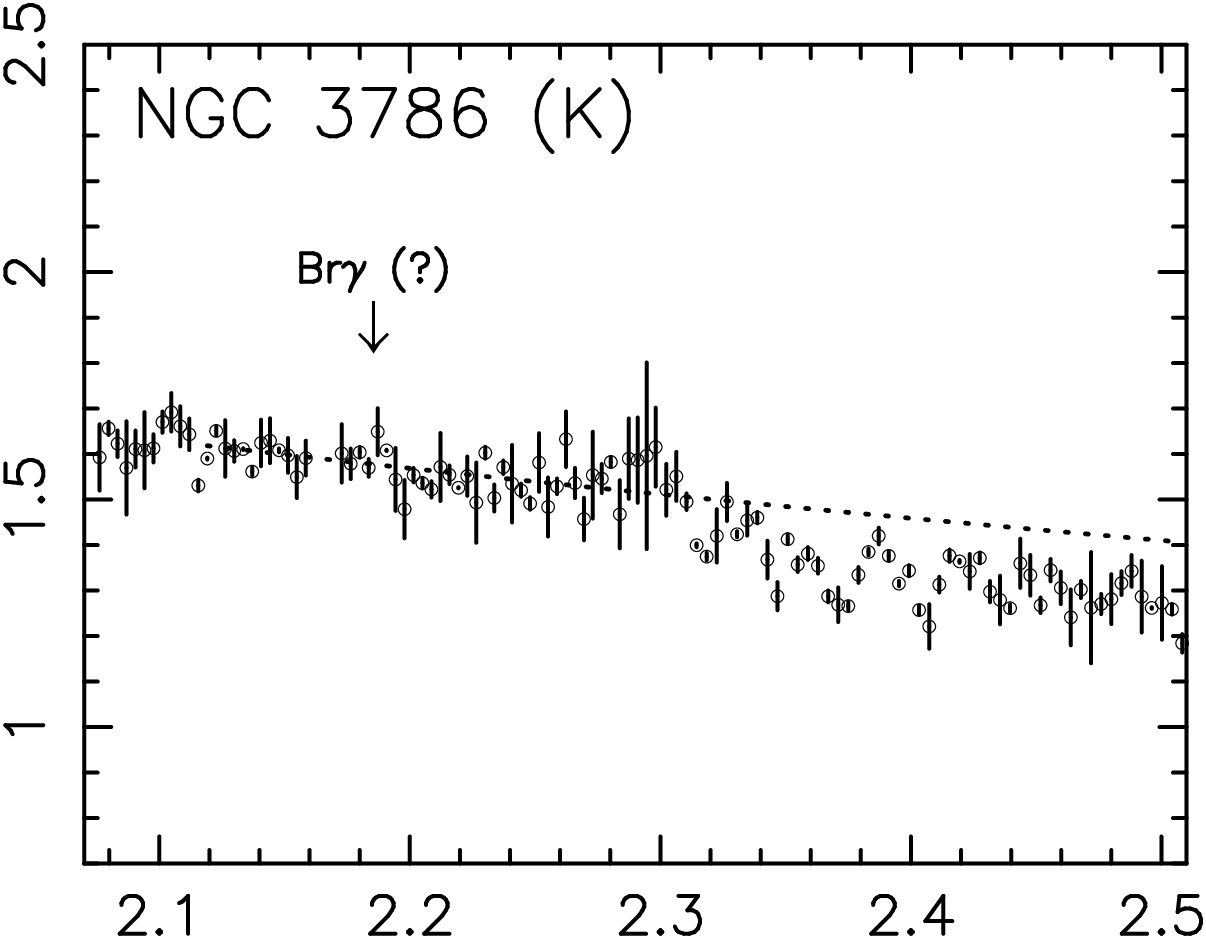} \hspace{0.3cm}
\includegraphics[angle=-90,scale=.35]{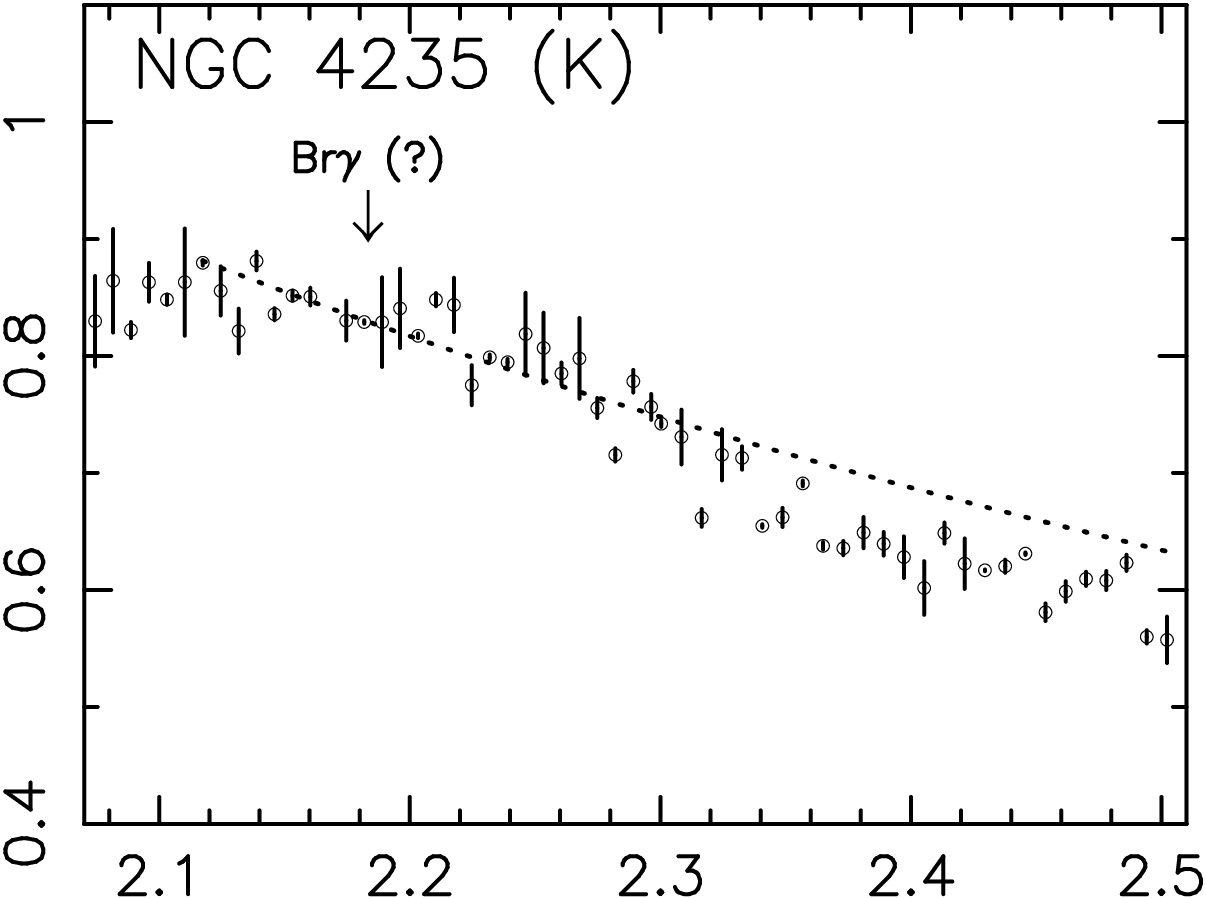} \\
\includegraphics[angle=-90,scale=.35]{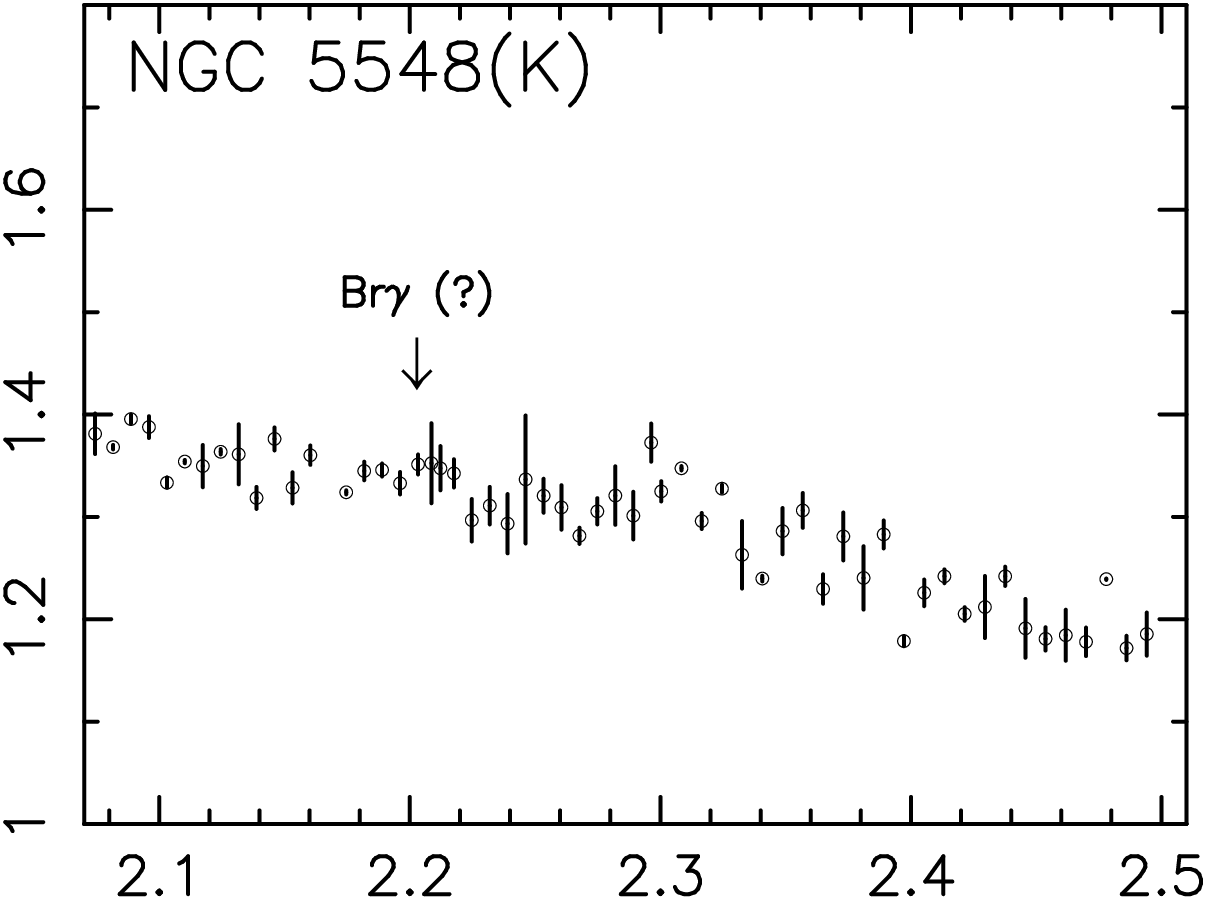} \hspace{0.3cm}
\includegraphics[angle=-90,scale=.35]{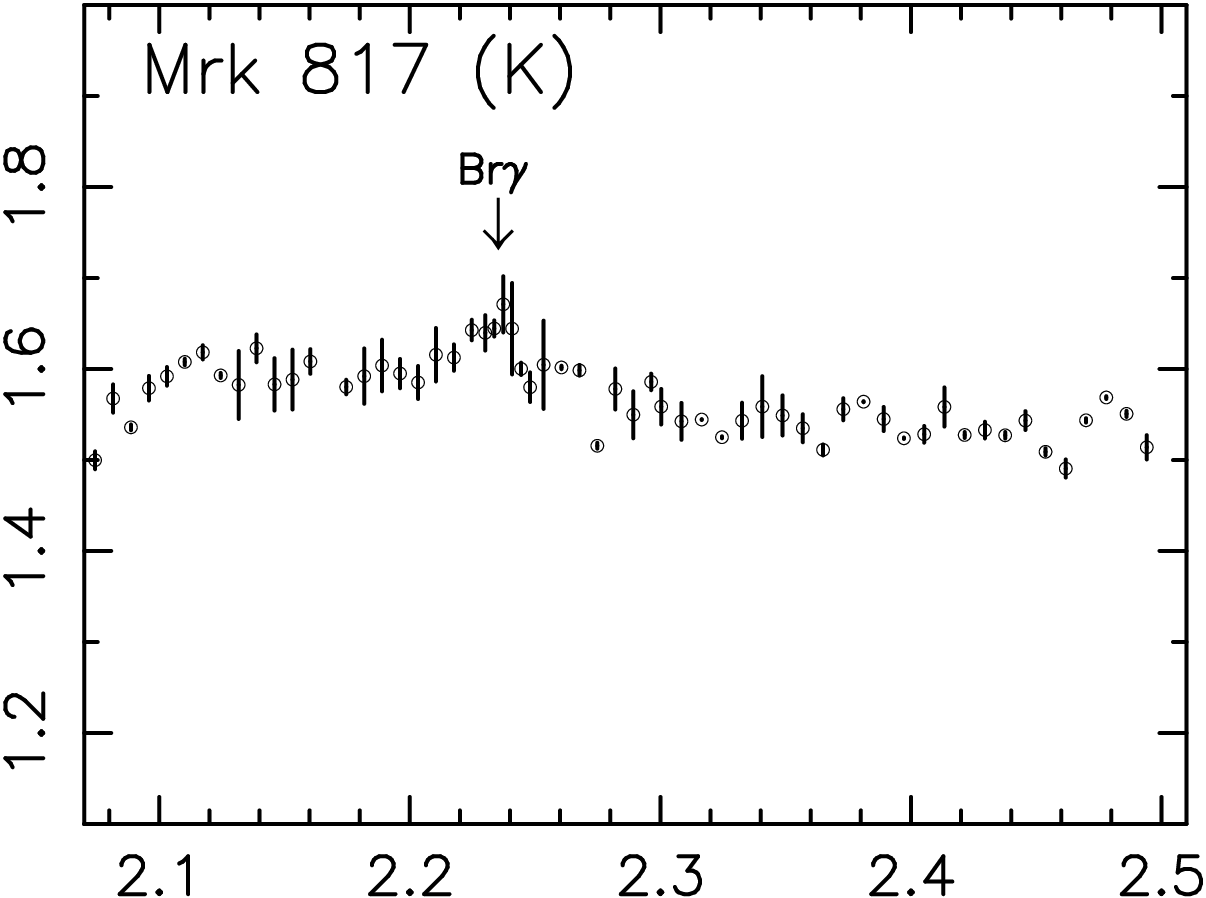} \\
\end{figure}

\clearpage

\begin{figure}
\includegraphics[angle=-90,scale=.35]{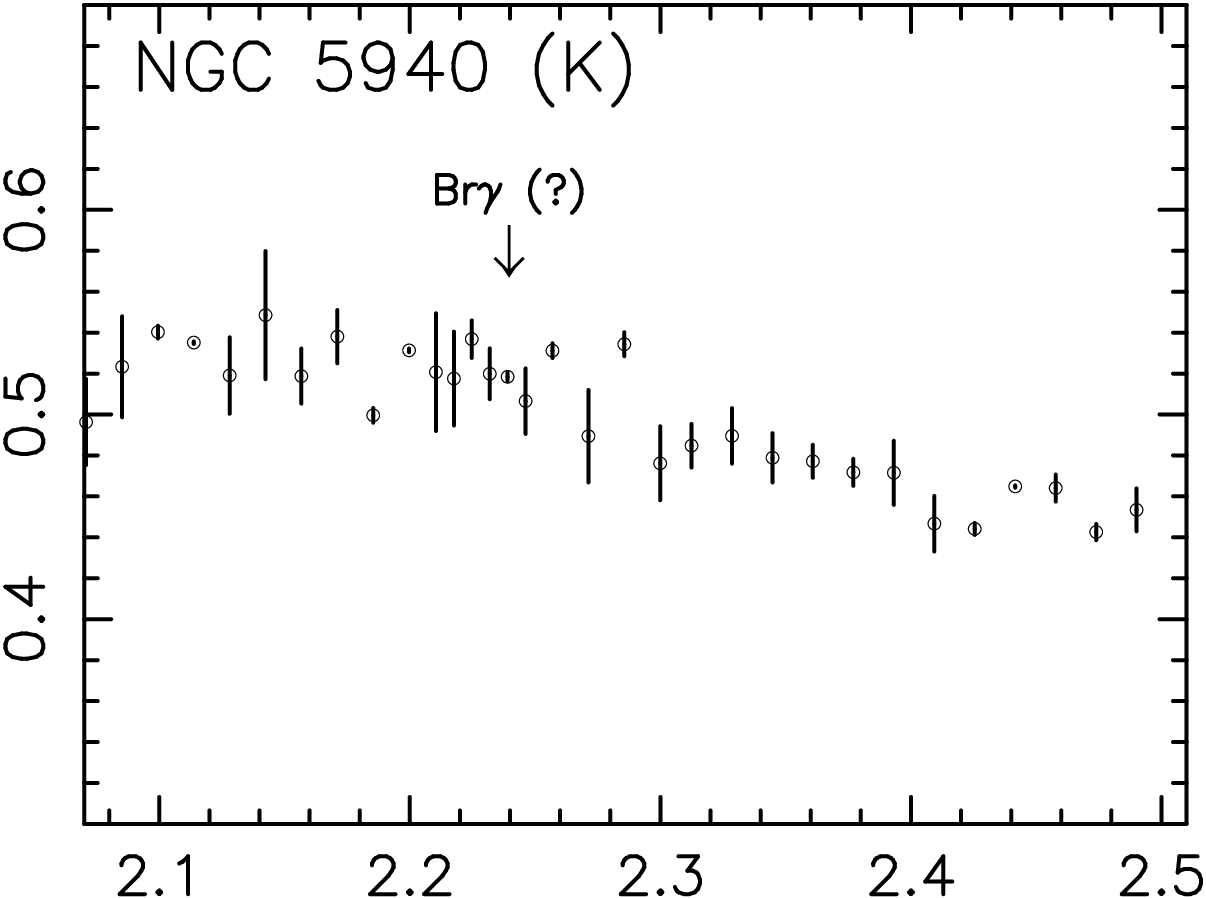} \hspace{0.3cm}
\includegraphics[angle=-90,scale=.35]{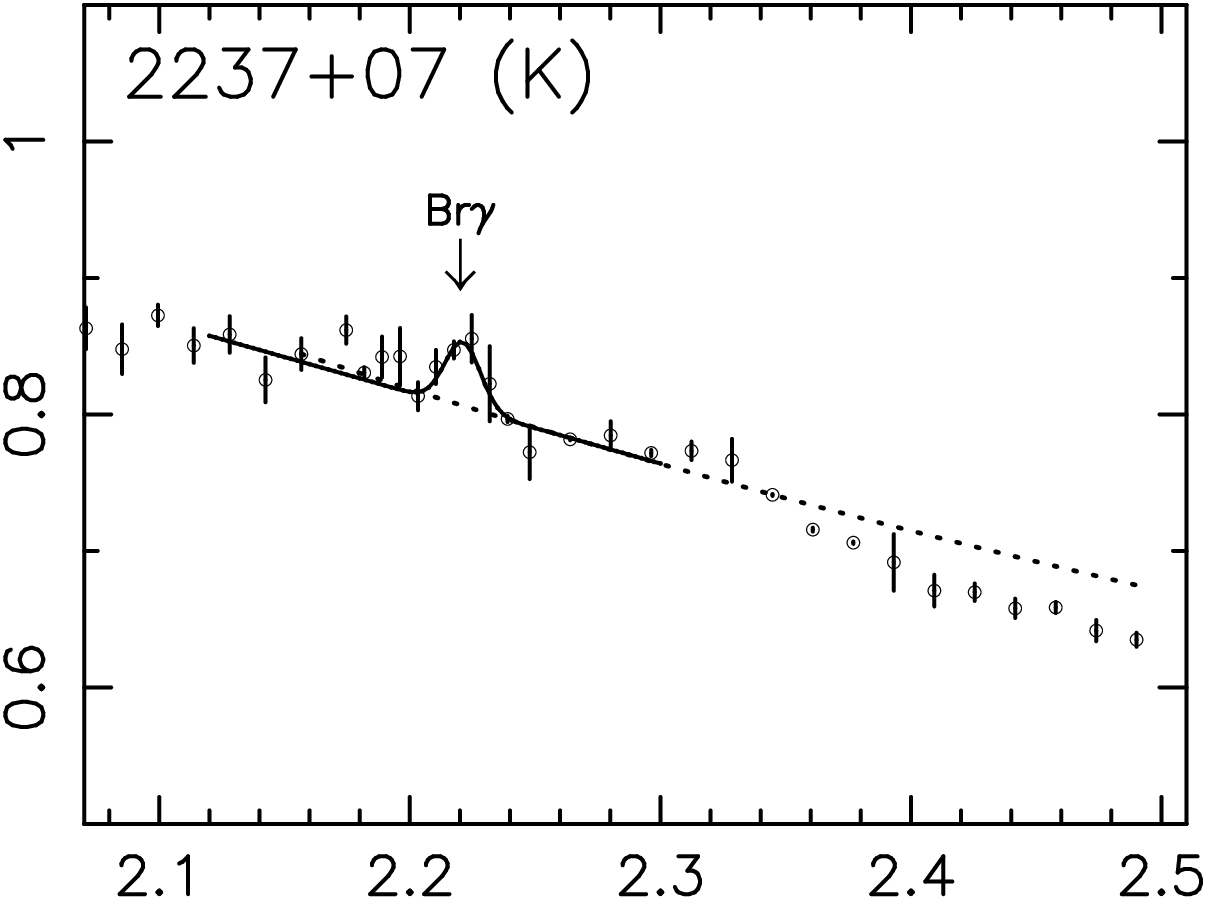} \\
\includegraphics[angle=-90,scale=.35]{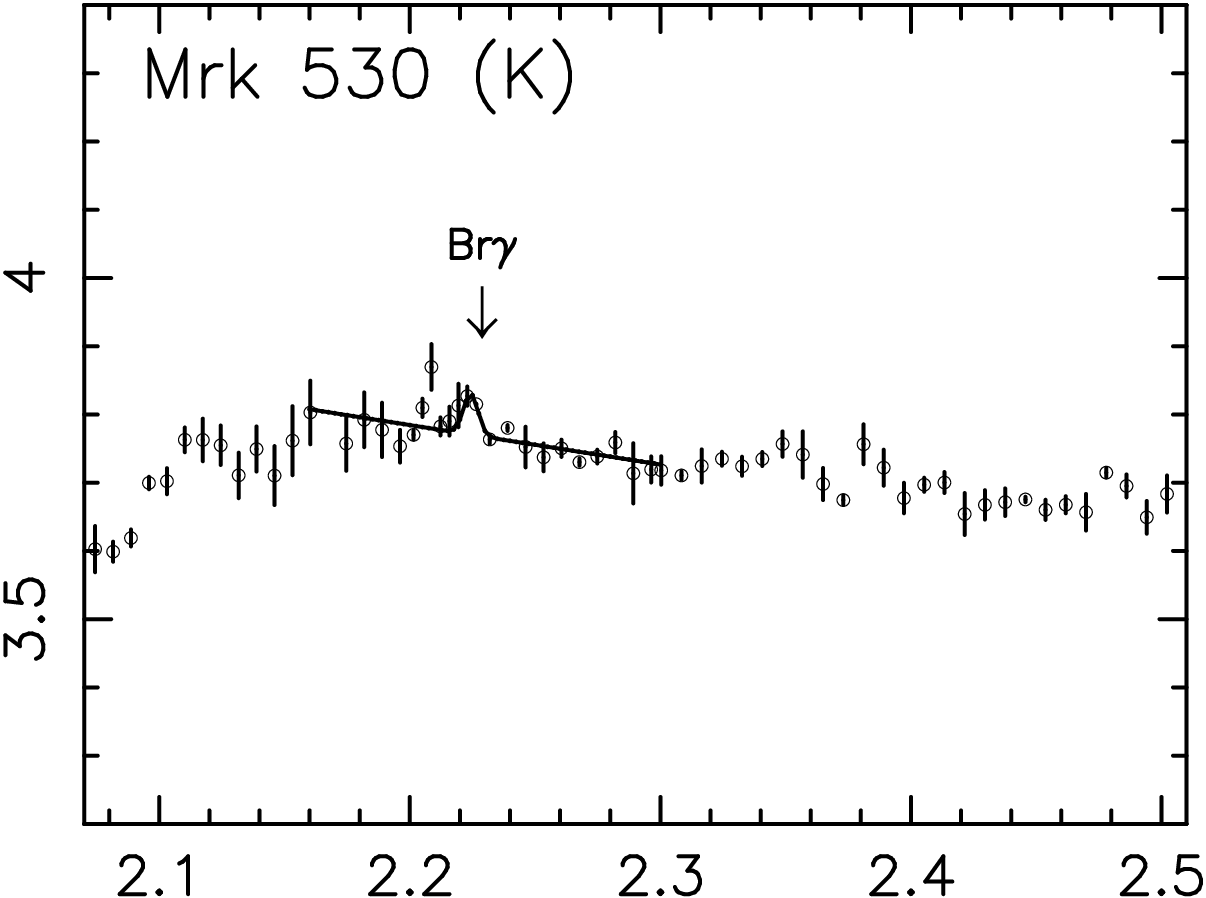} \hspace{0.3cm}
\includegraphics[angle=-90,scale=.35]{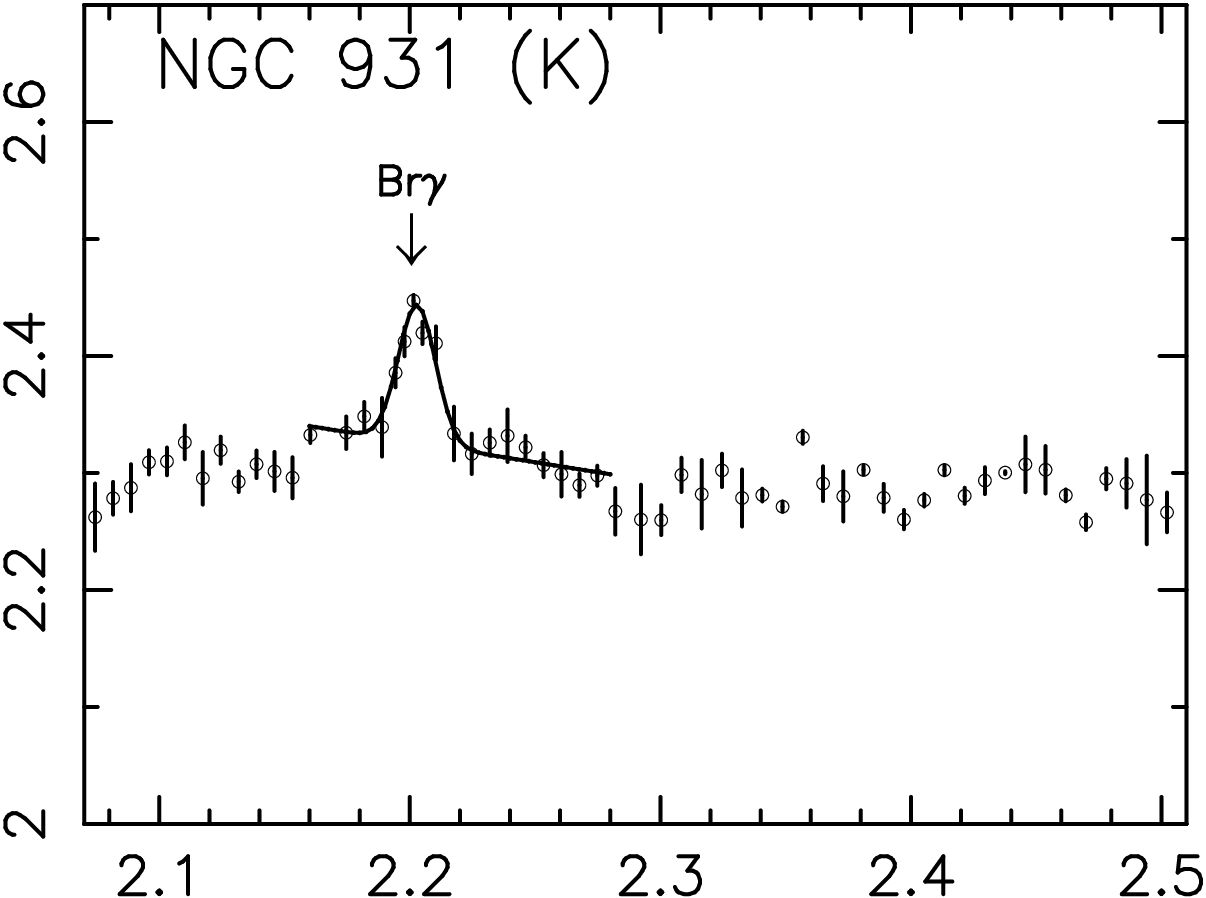} \\
\includegraphics[angle=-90,scale=.35]{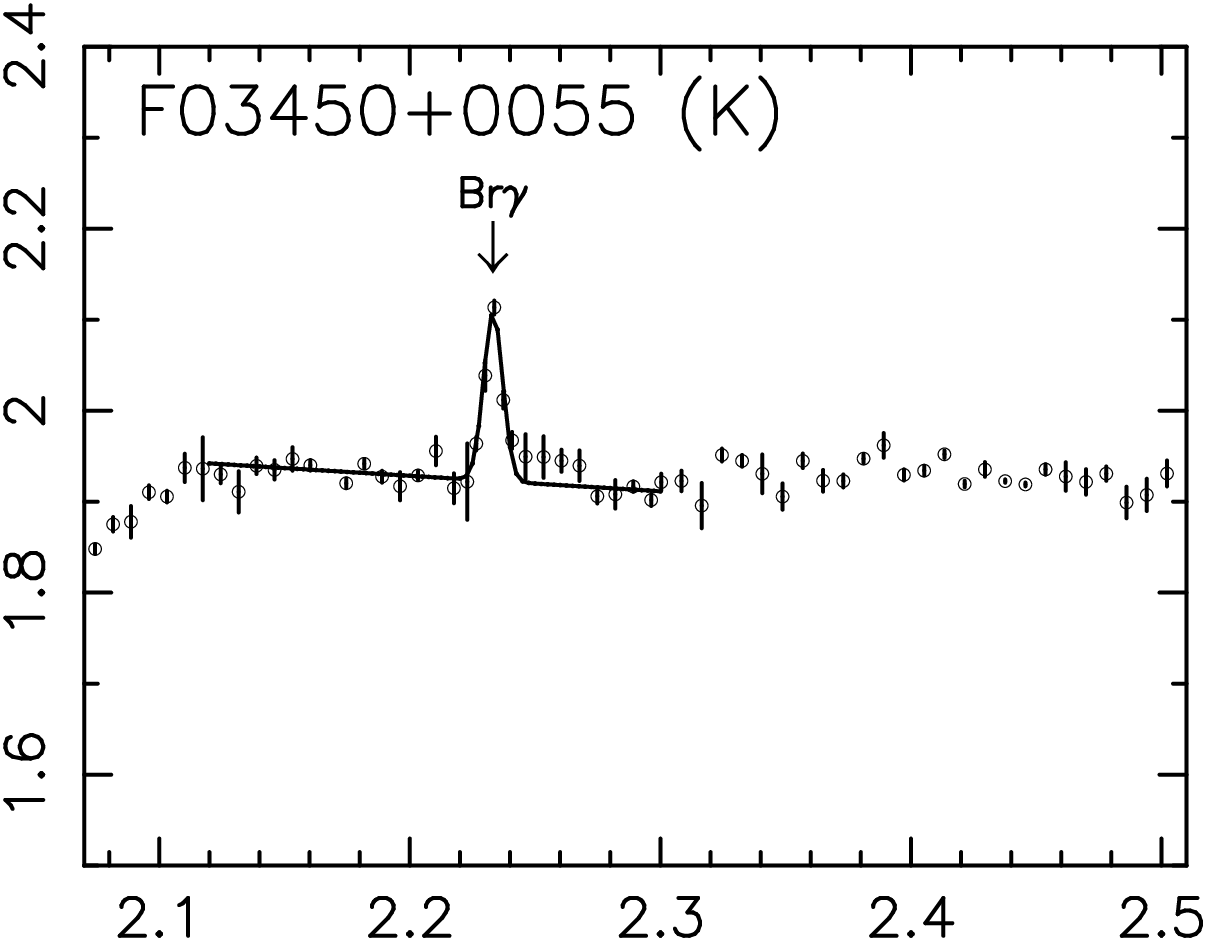} \hspace{0.3cm}
\includegraphics[angle=-90,scale=.35]{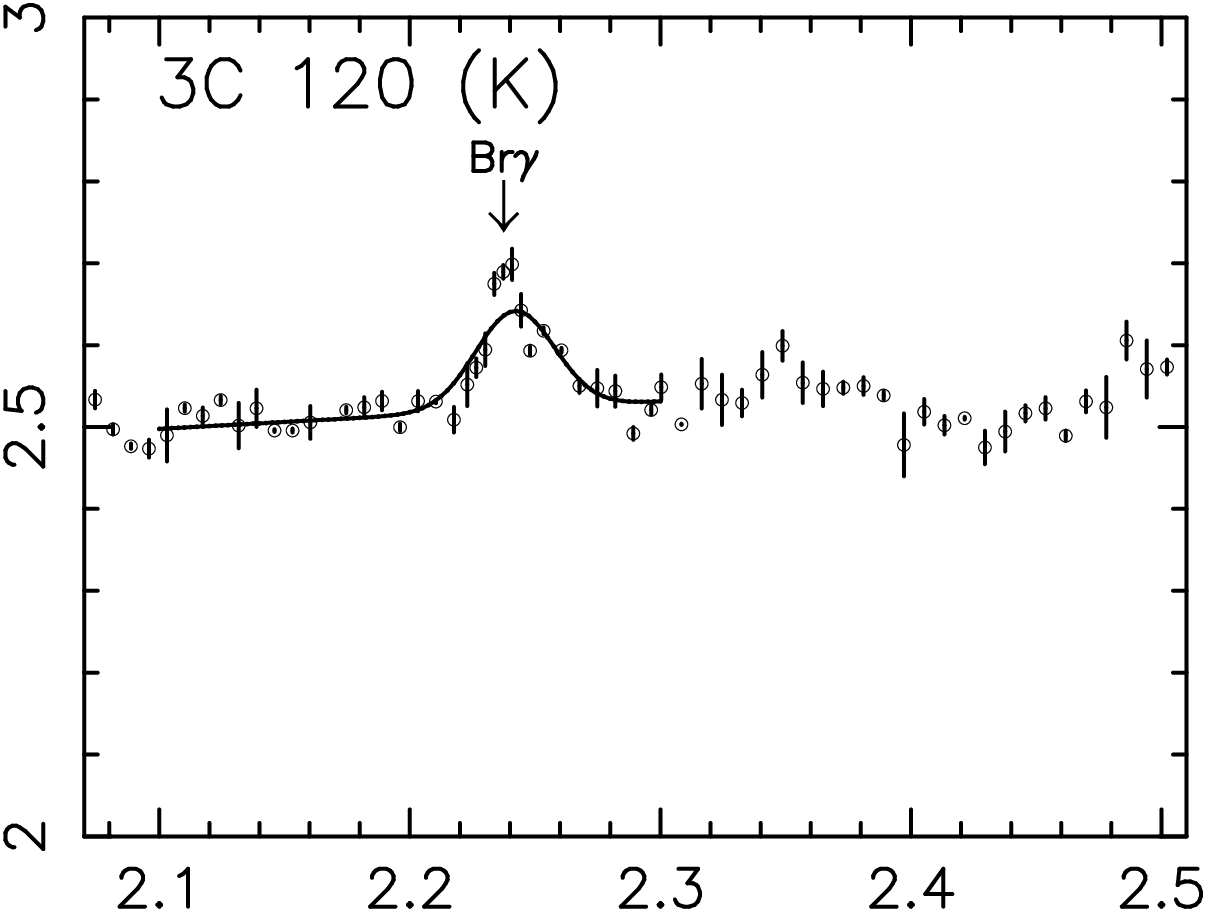} \\
\end{figure}

\clearpage

\begin{figure}
\includegraphics[angle=-90,scale=.35]{f2m.eps} \hspace{0.3cm}
\includegraphics[angle=-90,scale=.35]{f2n.eps} \\
\includegraphics[angle=-90,scale=.35]{f2o.eps} \hspace{0.3cm}
\includegraphics[angle=-90,scale=.35]{f2p.eps} \\
\includegraphics[angle=-90,scale=.35]{f2q.eps} \hspace{0.3cm}
\includegraphics[angle=-90,scale=.35]{f2r.eps}  \\
\end{figure}

\clearpage

\begin{figure}
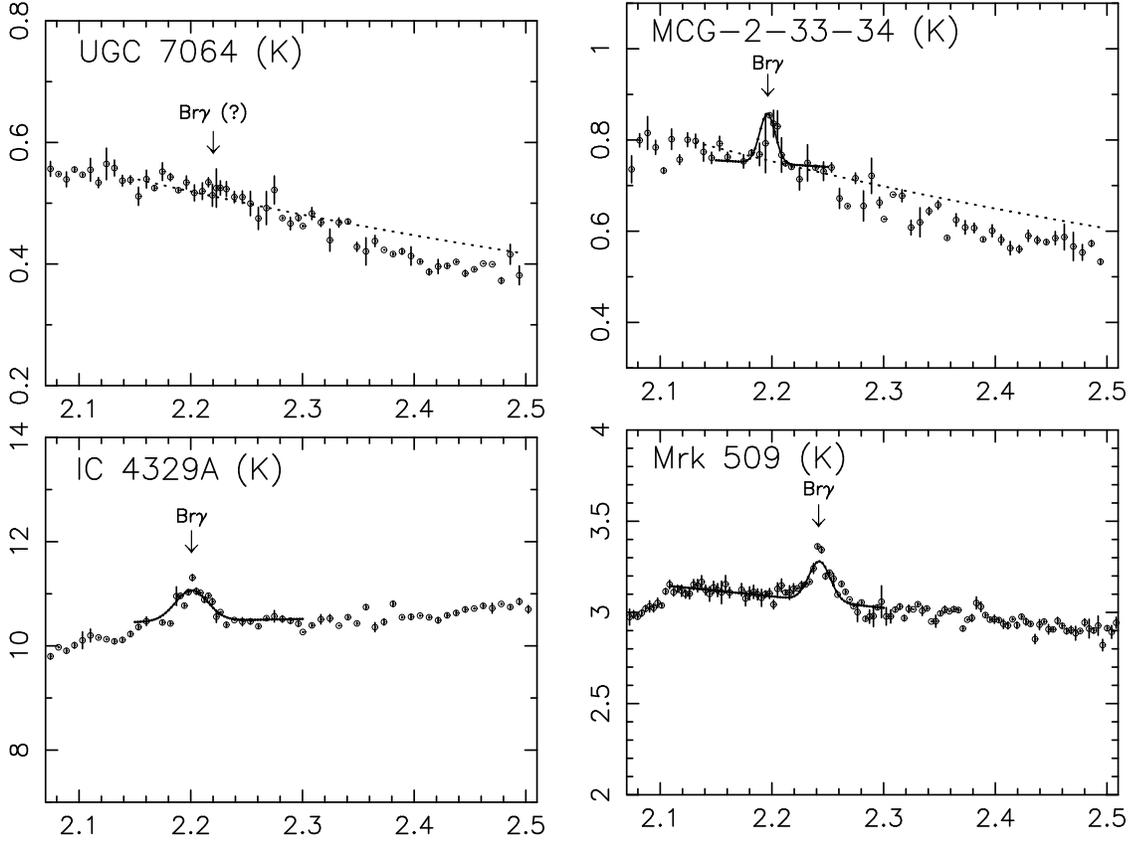

\includegraphics[angle=-90,scale=.35]{f2s.eps} \hspace{0.3cm}
\includegraphics[angle=-90,scale=.35]{f2t.eps} \\
\includegraphics[angle=-90,scale=.35]{f2u.eps}  \hspace{0.3cm}
\includegraphics[angle=-90,scale=.35]{f2v.eps} \\
\caption{Infrared $K$-band (2.07--2.5 $\mu$m) slit spectra of the 22
Seyfert 1 nuclei.
The abscissa and ordinate are the observed wavelength in $\mu$m and
F$_{\lambda}$ in 10$^{-14}$ W m$^{-2}$ $\mu$m$^{-1}$, respectively.
For sources with clear Br$\gamma$ emission lines, adopted Gaussian fits
are overplotted as solid lines.
For sources with signatures of CO absorption features, adopted continuum
levels, with which to measure the absorption strengths, are shown as
dashed lines.
}
\end{figure}

\clearpage

\begin{figure}
\includegraphics[angle=0,scale=.45]{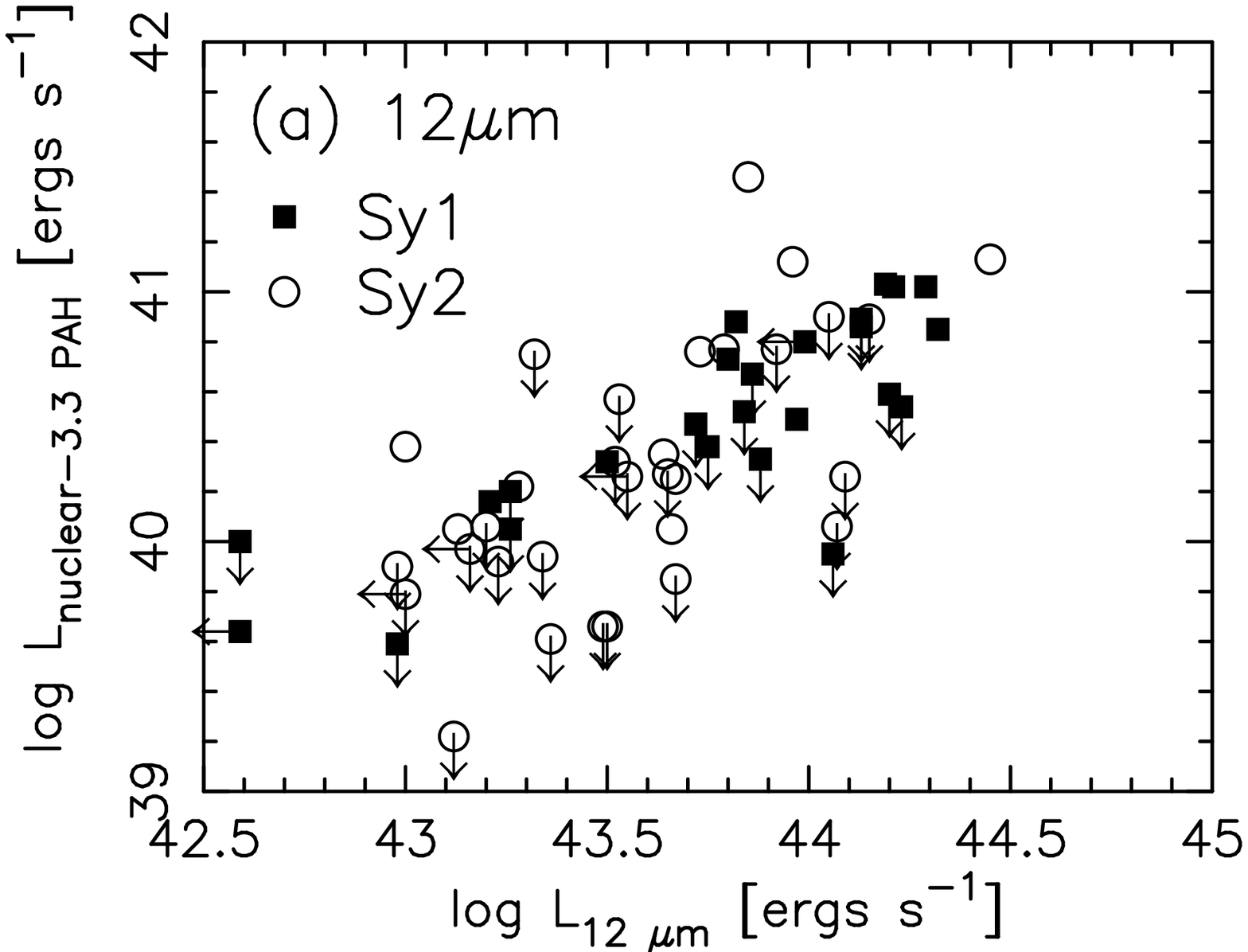}
\includegraphics[angle=0,scale=.45]{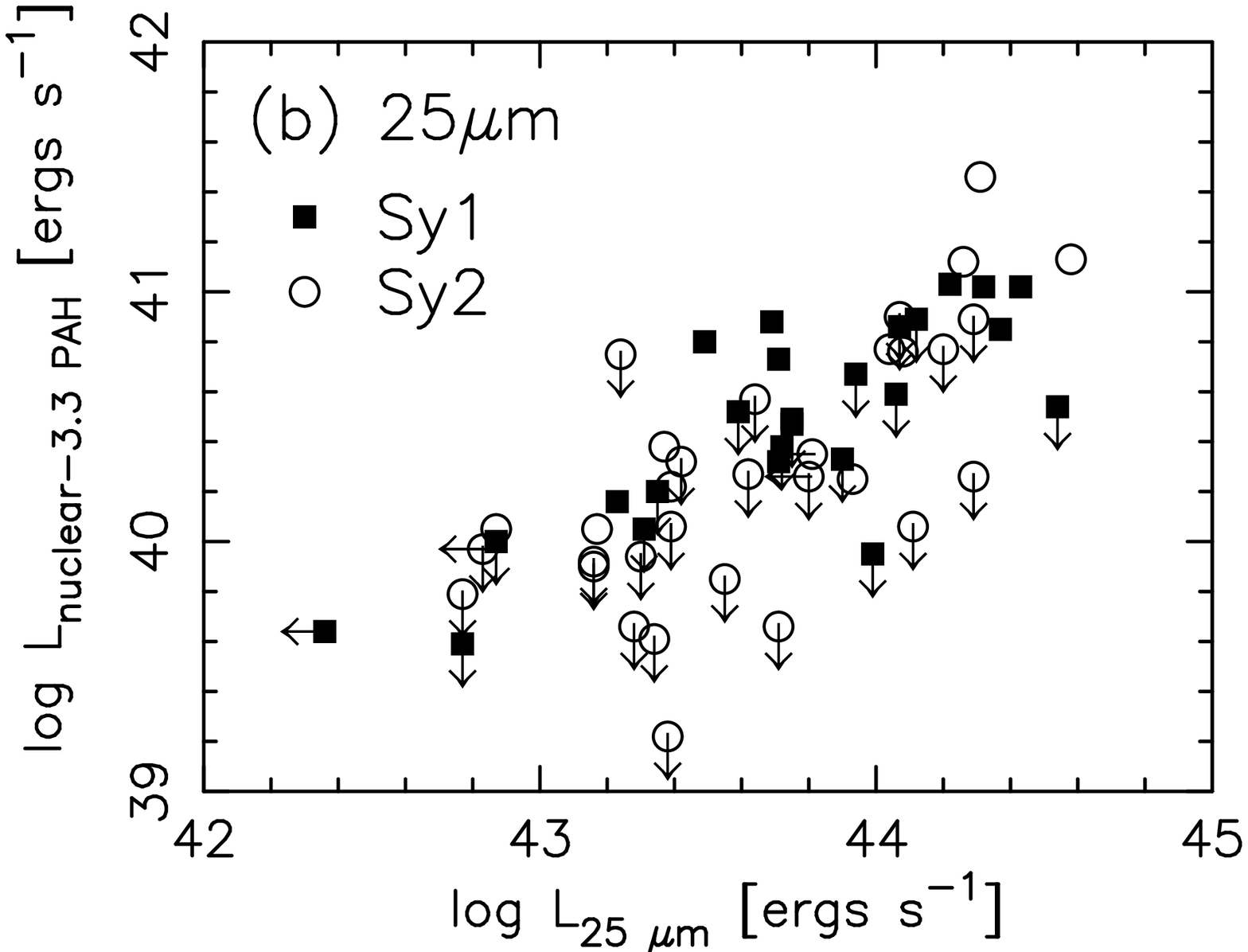}
\includegraphics[angle=0,scale=.45]{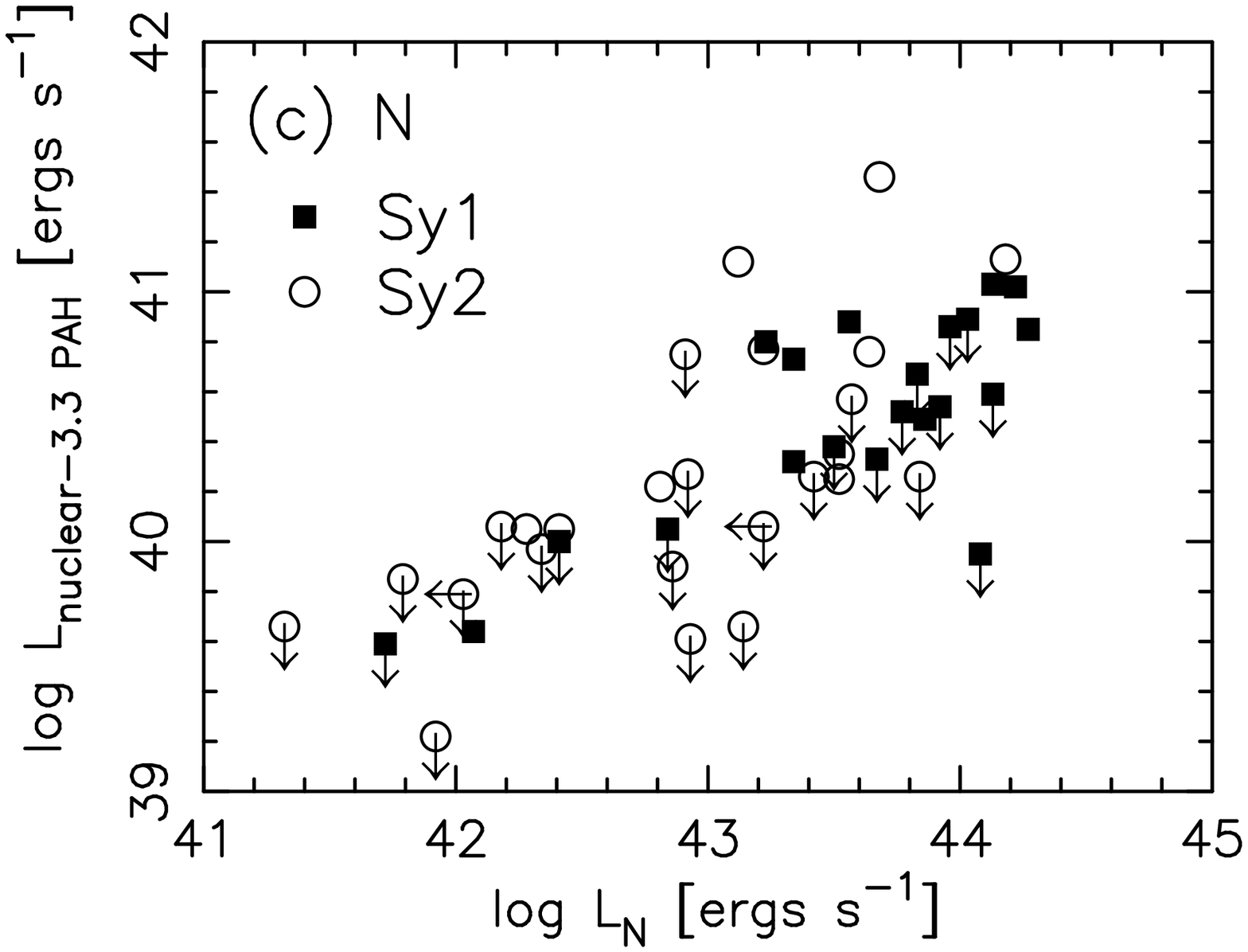}
\includegraphics[angle=0,scale=.45]{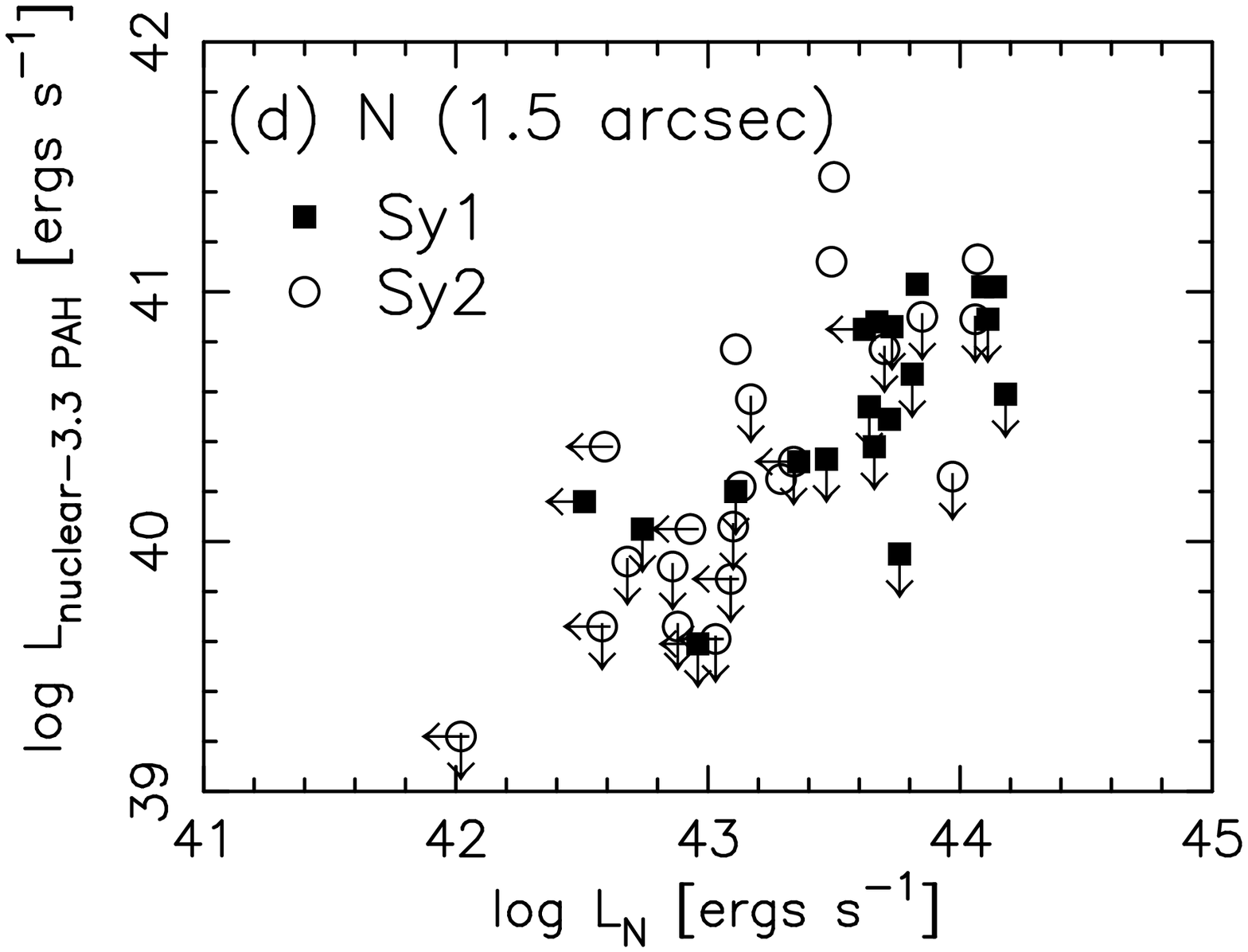}
\end{figure}
\clearpage
\begin{figure}
\includegraphics[angle=0,scale=.45]{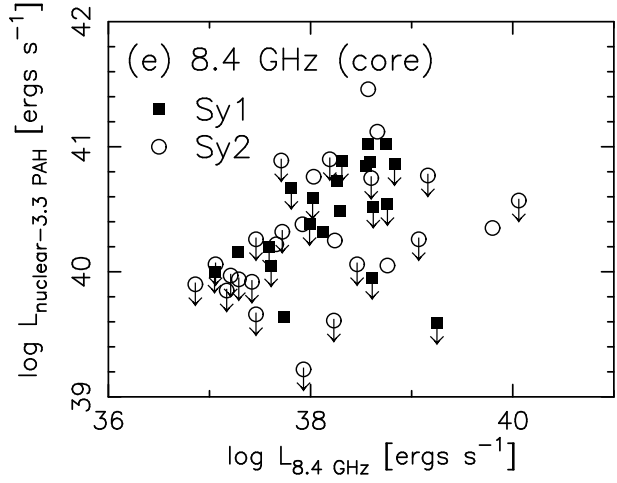} \\
\caption{
{\it (a)}: Comparison of the {\it IRAS} 12 $\mu$m luminosity,
defined as $\nu$L$_{\nu}$ (abscissa), and nuclear 3.3 $\mu$m PAH
emission luminosity detected inside our slit spectra (ordinate).
Filled squares are data for Seyfert 1s, and open circles are
for Seyfert 2s taken from \citet{ima03}.
{\it (b)}: Same as (a), but the abscissa is {\it IRAS} 25 $\mu$m luminosity.
{\it (c)}: The abscissa is $N$-band (10.6 $\mu$m)
luminosity measured with ground-based aperture ($>$4$''$) photometry,
using a single pixel detector.
{\it (d)}: The abscissa is $N$-band (10.6 $\mu$m)
luminosity measured with a ground-based two-dimensional camera with a
1$\farcs$5 aperture \citep{gor04}.
{\it (e)}: The abscissa is core radio luminosity at 8.4 GHz measured
with 0$\farcs$25 spatial resolution \citep{kuk95,the00}.
In all the plots, if Seyfert 2s have intrinsically stronger nuclear
starbursts, with respect to AGN powers, than Seyfert 1s, their location
should be to the upper-left of the Seyfert 1' distribution.
}
\end{figure}

\clearpage

\begin{figure}
\includegraphics[angle=0,scale=.45]{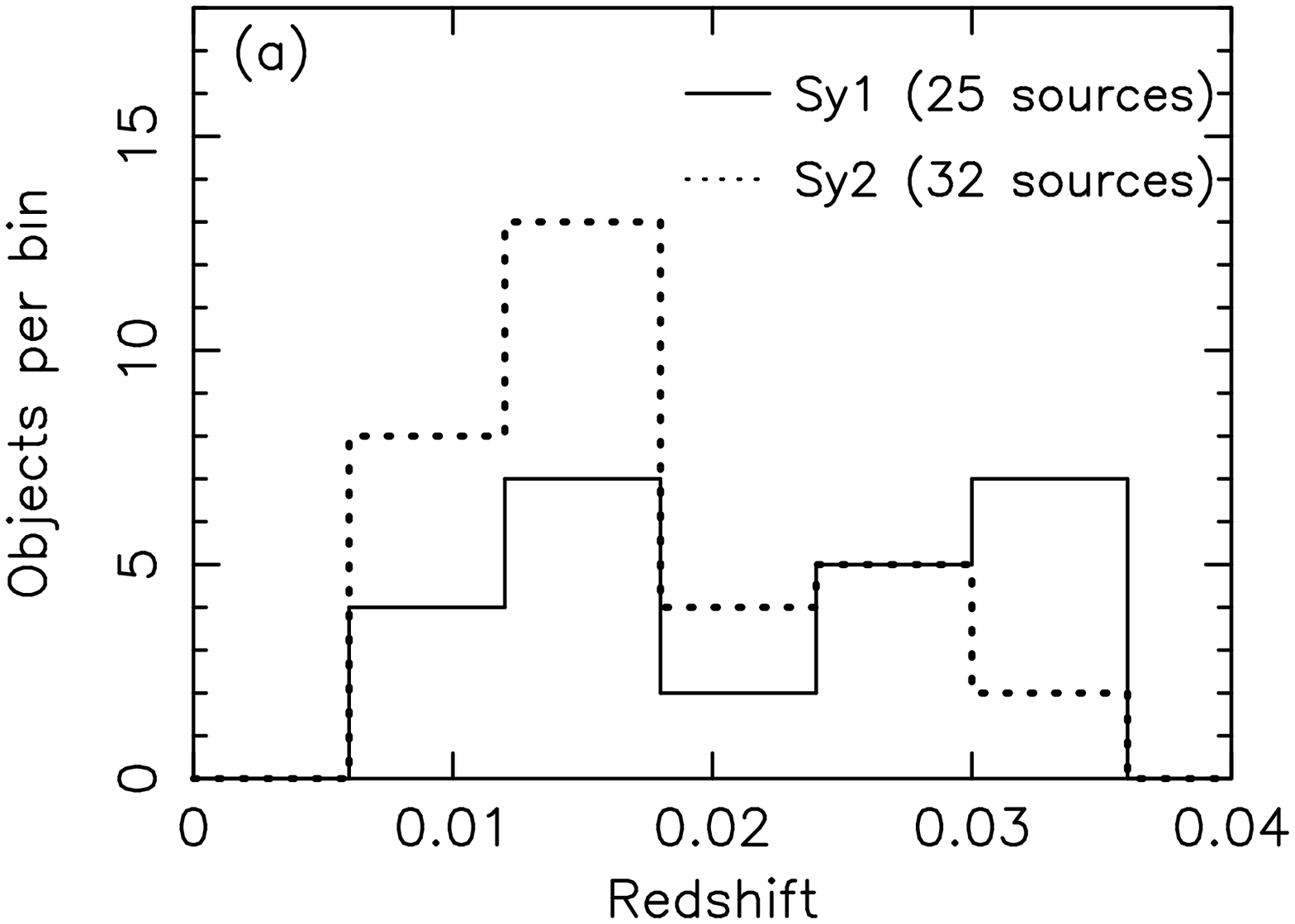}
\includegraphics[angle=0,scale=.45]{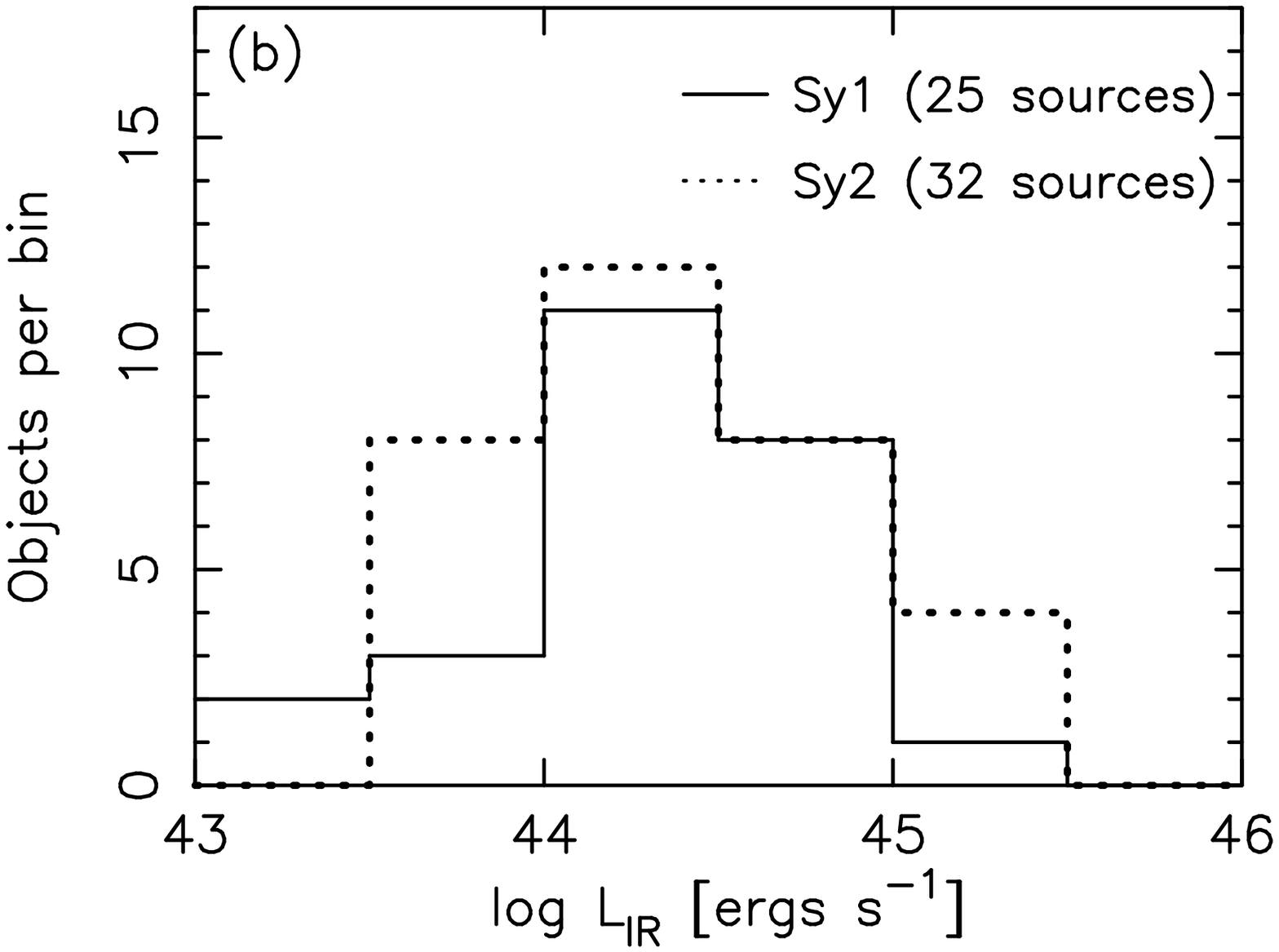}
\caption{
{\it (a)}: Histogram of redshifts.
Solid lines are data for Seyfert 1s, and dashed lines are for Seyfert 2s
studied by \citet{ima03}. 
{\it (b)}: Histogram of the infrared luminosities.
For sources that have upper limits in some {\it IRAS} bands,
we assume that the actual flux is the upper limit.
For these sources, even though the lower values of infrared
luminosities (Table 1) are adopted, there is no systematic difference in
the distribution of the infrared luminosities between Seyfert 1s and 2s.  
}
\end{figure}

\clearpage

\begin{figure}
\includegraphics[angle=0,scale=.45]{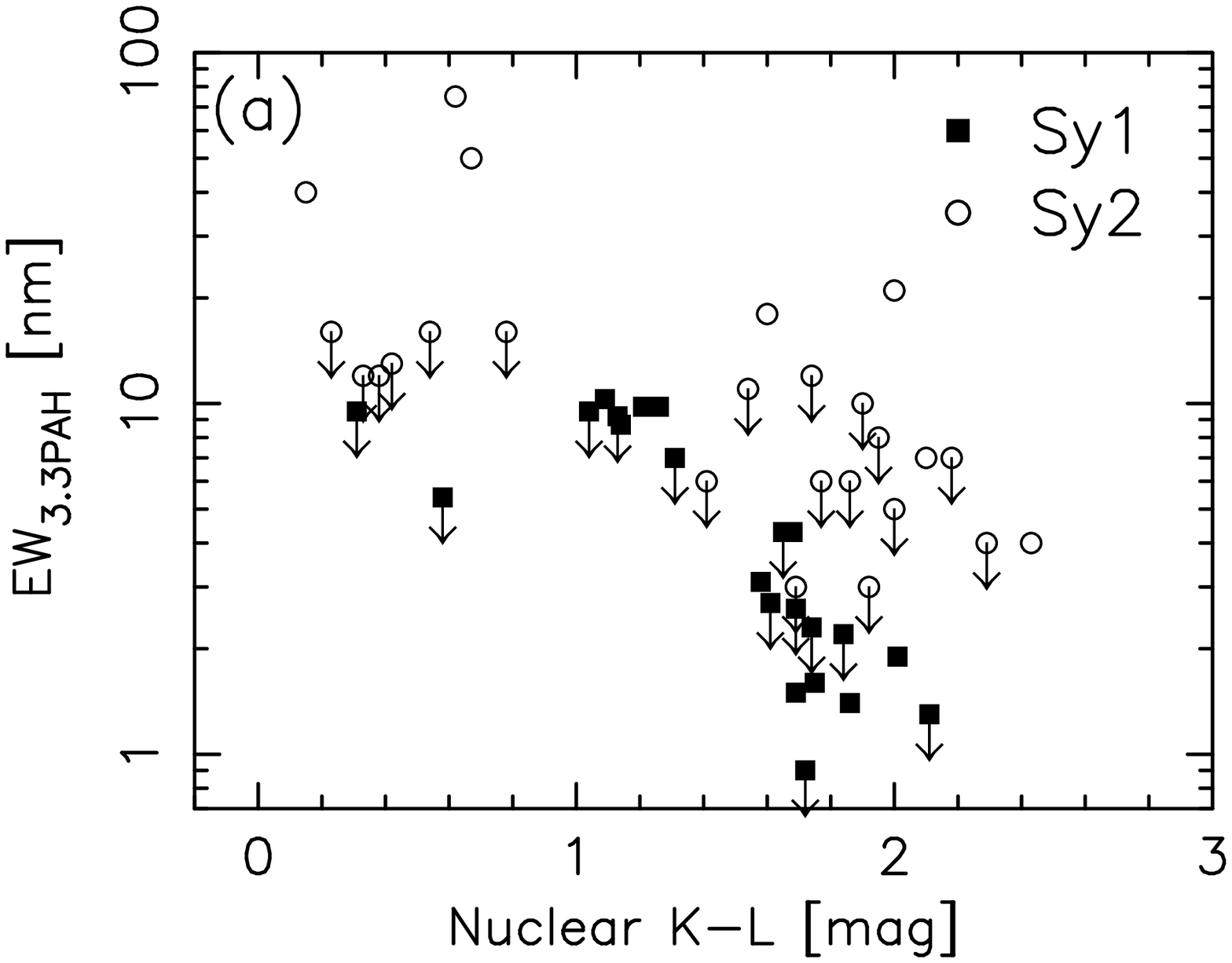}
\includegraphics[angle=0,scale=.45]{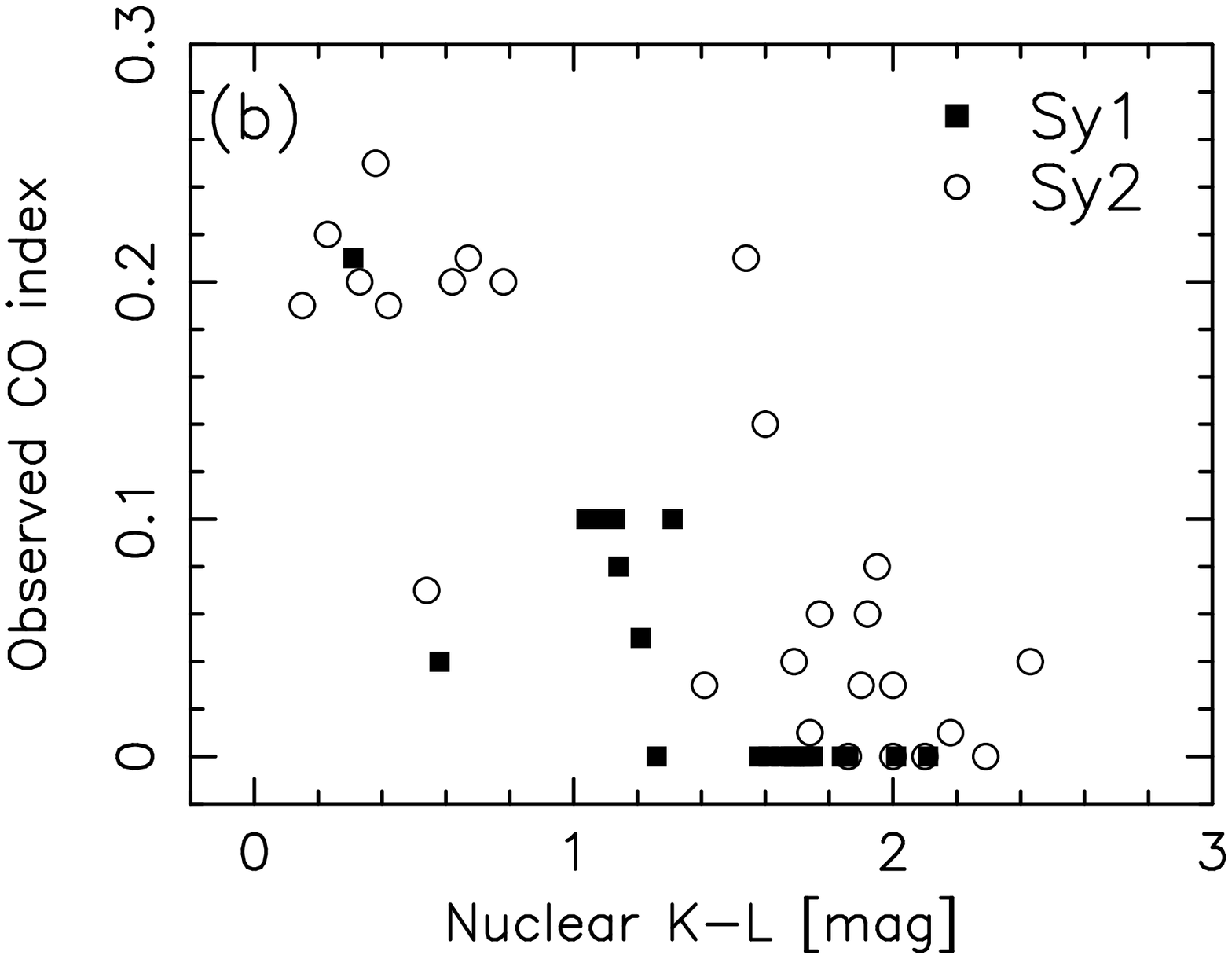}
\caption{
{\it (a):} Nuclear $K - L$ magnitude (abscissa) and observed rest-frame
equivalent width of the 3.3 $\mu$m PAH emission feature (ordinate).
Filled squares are data for Seyfert 1s, and open circles are
data for Seyfert 2s adopted from \citet{ima03}.
{\it (b):} The ordinate is the observed spectroscopic CO index.
Filled squares are data for Seyfert 1s, and open circles are
for Seyfert 2s taken from \citet{ia04}.
}
\end{figure}

\clearpage

\begin{figure}
\includegraphics[angle=0,scale=.45]{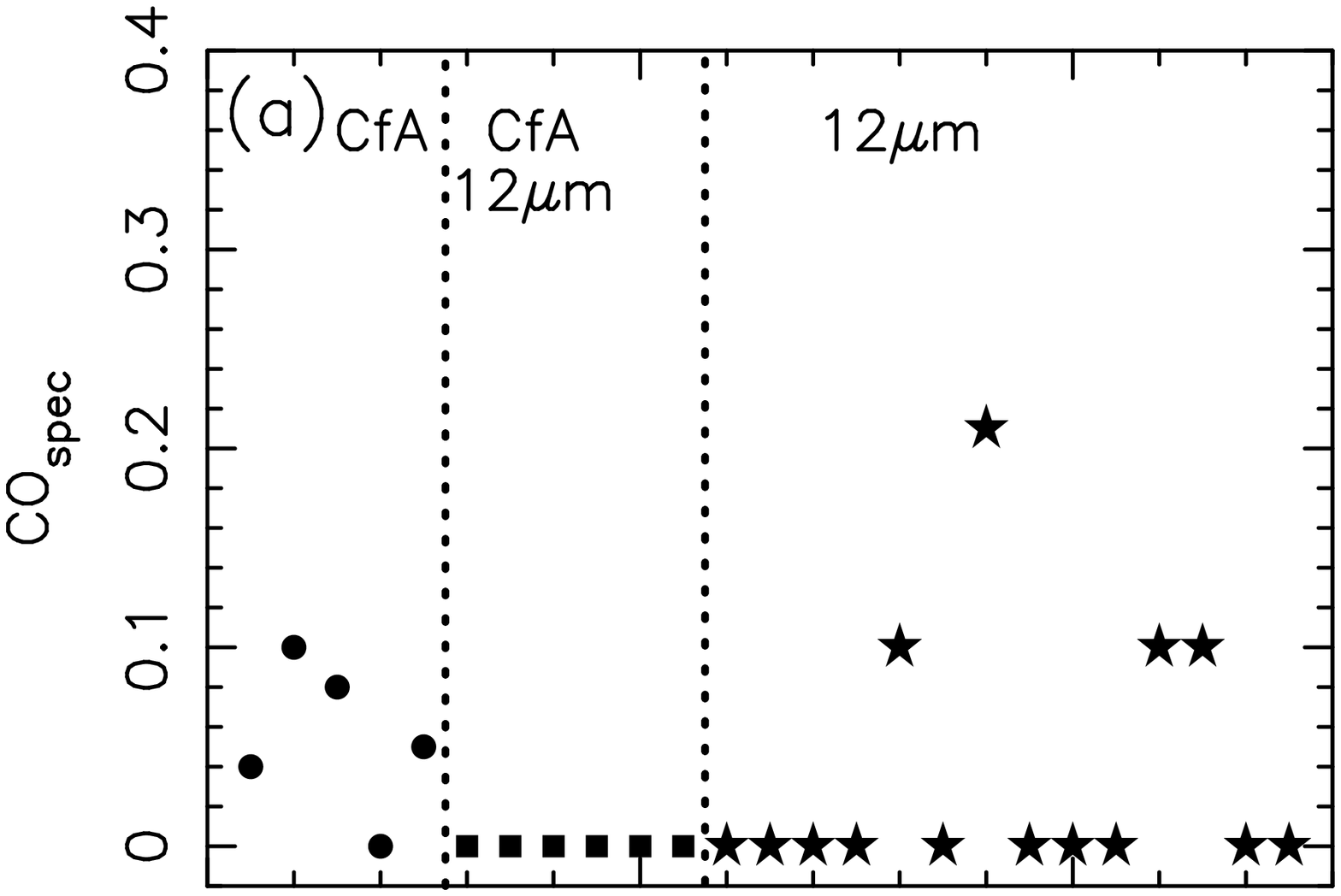}
\includegraphics[angle=0,scale=.45]{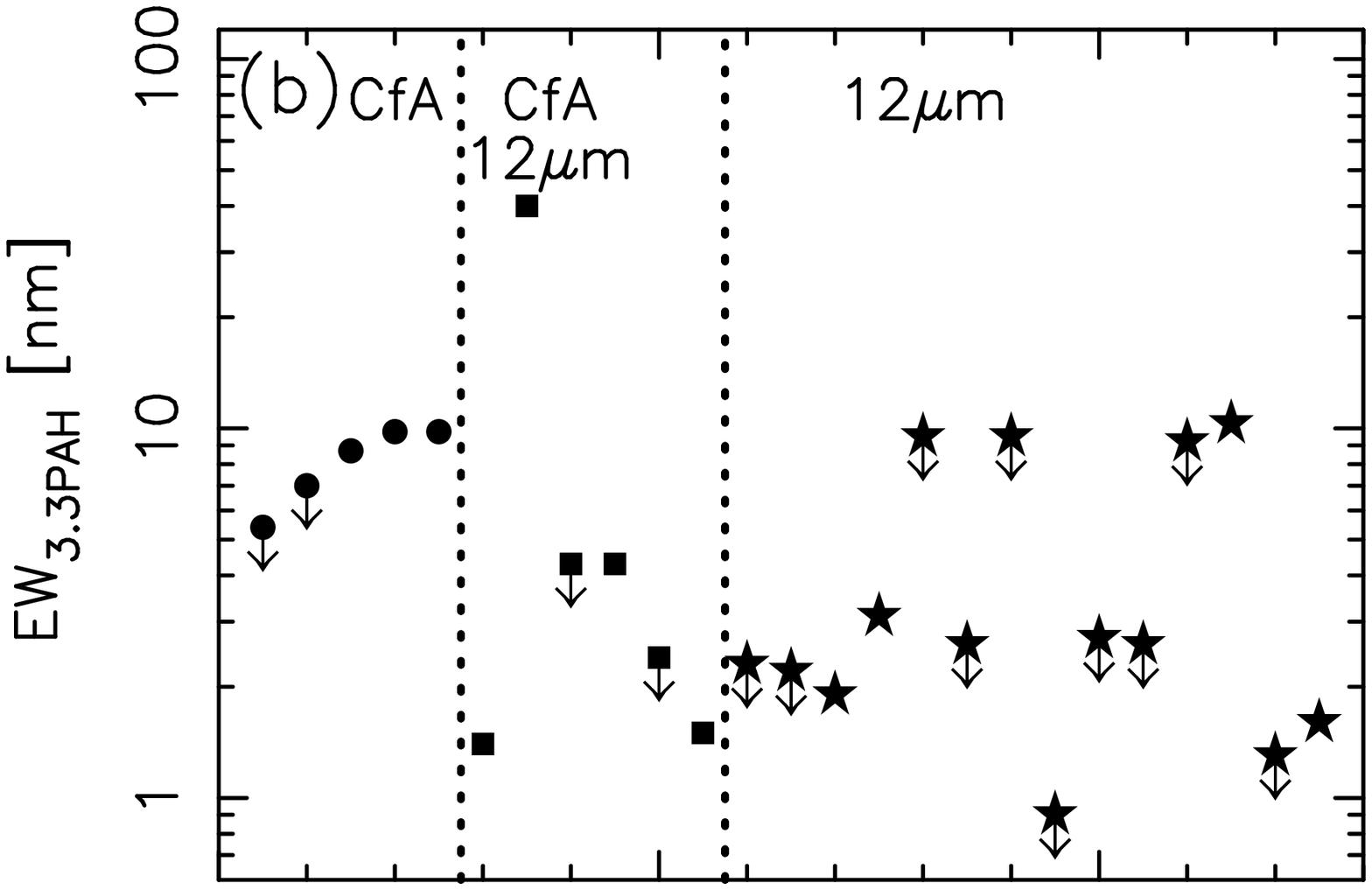}
\includegraphics[angle=0,scale=.45]{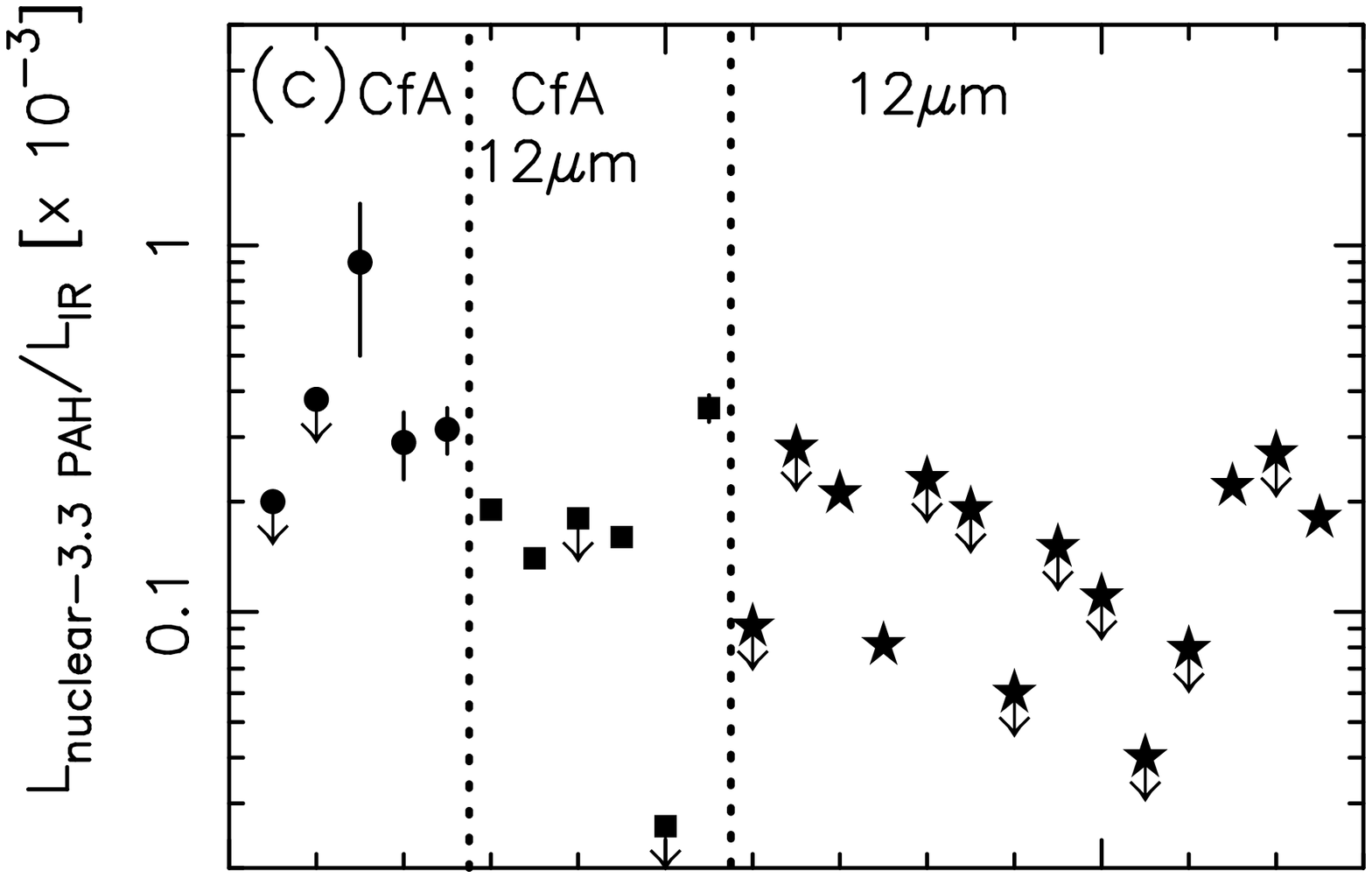}
\caption{
{\it (a):} Distribution of the observed spectroscopic CO index.
Solid circles: CfA Seyfert 1s. Solid stars: 12 $\mu$m Seyfert 1s.
Six Seyfert 1s listed in both samples are plotted as solid squares.
{\it (b)}: Rest-frame equivalent width of the 3.3 $\mu$m PAH
emission feature.
{\it (c)}: Nuclear 3.3 $\mu$m PAH to infrared luminosity ratio.
}
\end{figure}

\end{document}